\documentclass[a4paper,12pt]{article}
\usepackage[T1]{fontenc}
\usepackage{fullpage}
\usepackage{graphicx}
\usepackage{geometry}
\usepackage{float}
\usepackage{amsmath}
\usepackage{amssymb}
\usepackage{xcolor}
\usepackage{colortbl}
\usepackage{cancel}
\usepackage[normalem]{ulem}
\usepackage[left]{lineno}
\usepackage{blindtext}

\usepackage{hyperref}
\usepackage{centernot}
\usepackage{amsthm}
\usepackage{tikz}

\usepackage[lofdepth,lotdepth]{subfig}
\usepackage{caption}
\usepackage[affil-it, auth-lg]{authblk}
\usepackage{cite}

\theoremstyle{remark}

\newcommand{\be}{\begin{equation}}
\newcommand{\ee}{\end{equation}}
\newcommand{\nc}{\newcommand}
\nc{\ssi}{\sum_{i=1}^n}
\nc{\ssj}{\sum_{j=1}^n}
\nc{\ssk}{\sum_{k=1}^n}

\definecolor{viridis2}{RGB}{33,145,140}
\definecolor{viridis3}{RGB}{94,201,98}
\definecolor{viridis4}{RGB}{173,220,51}
\definecolor{viridis5}{RGB}{253,231,37}

\title{Symmetries of weighted networks: \\ weight approximation method \\ and its application to food webs}
\author[1]{Mateusz Iskrzy\'nski}
\author[2]{Julia Korol}
\author[2,3,4,*]{Aleksandra Puchalska}
\affil[1]{\scriptsize{Polish Academy of Sciences, Systems Research Institute, Warsaw, Poland}}
\affil[2]{University of Warsaw, Institute of Applied Mathematics and Mechanics, Warsaw, Poland}
\affil[3]{University of Ljubljana, Faculty of Mathematics and Physics, Ljubljana, Slovenia}
\affil[4]{Institute of Mathematics, Physics and Mechanics, Ljubljana, Slovenia}
\affil[*]{apuchalska@mimuw.edu.pl}

\begin{document}

\maketitle

\begin{abstract}
Graph symmetries identify structural regularities and reduce the computational complexity of network analysis. In weighted graphs, however, exact automorphisms are rare because real-valued weights seldom coincide. 
We introduce a general framework for detecting approximate symmetries by aggregating weights into discrete categories, generating a sequence of coarser graphs on which classical automorphism analysis applies. The approximation path is fully configurable, based on interaction magnitudes, and can be matched to the empirical weight distribution.

Applied to 250 empirical food webs using logarithmic aggregation, the method reveals that automorphisms emerge even at low approximation levels and almost always form small orbits. Orbit sizes rarely exceed two or three vertices, reflecting the combinatorial fragility of larger symmetric sets. Even so, symmetric vertices occupy diverse  structural positions in the network and high connectivity does not imply asymmetry. The observation of just local permutations confirms the conclusions of trophic species and niche analysis. A case study demonstrates that automorphisms can also recover latent ecological structure. The minimal aggregation level at which two vertices become substitutable provides a quantitative measure of role similarity. The framework offers a principled, automorphism-based approach for quantifying similarity and redundancy in weighted complex networks.
\end{abstract}

\section{Introduction}

Understanding which vertices play similar roles is a central problem in network science. The first indication of similarity is given by having the same value of a graph measure, e.g. vertex degree. Several concepts explore similarity understood in a much stronger sense, and from complementary angles. Structural equivalence~\cite{LORRAIN} requires vertices to have either identical or similar~\cite{Granovetter_1973, HollandLeinhardt_1975, Burt_1976, EverettBorgatti_1991} neighbourhoods. Regular equivalence~\cite{Everett_1994} groups vertices that connect to others with similar roles, and automorphic equivalence identifies vertices symmetric under a graph automorphism, a permutation of vertices that preserves the structure of the entire network~\cite{EVERETT1985353, MacArthur20083525, Garlaschelli_review, SanchezGarcia2020}. Structurally equivalent vertices are also symmetric under swapping as all their connections are identical, while any automorphically equivalent vertices are, by definition, regularly equivalent. These concepts thus form a hierarchy of increasingly restrictive relationships.

In this work, we introduce a framework for detecting and quantifying symmetries (automorphisms) in weighted networks. Empirical networks often contain real-valued edge weights estimated with uncertainty, which makes exact automorphisms rare. We propose an iterative \textit{weight aggregation} procedure that discretises edge values, enabling standard symmetry analysis on a nested sequence of graphs with increasingly coarser weight resolution. This approach allows the detection of functionally similar vertices even when exact numerical symmetry is absent. We demonstrate its use on a dataset of 250 empirical weighted food webs that portray biomass flows in natural ecosystems and evaluate three mutually complementary measures of network-level symmetry. 

These three forms of equivalence--structural, regular, and automorphic--capture complementary notions of similarity. Early quantitative studies of \textit{approximate structural equivalence} compared adjacency patterns through correlation or distance measures~\cite{Granovetter_1973, HollandLeinhardt_1975, Burt_1976, EverettBorgatti_1991}, while the concept of \textit{regular equivalence}~\cite{WhiteReitz1983, Everett_1994} generalised this idea to roles: vertices are equivalent if they connect to others that are themselves equivalent, even if their specific neighbors differ. These developments were later integrated into blockmodeling frameworks that identify near-equivalence under controlled error tolerance~\cite{BORGATTI1993361, Doreian2005}. A recent parallel approach extends these principles to weighted and noisy networks by allowing small edge-weight tolerances~\cite{Squillace_2025}. Across this spectrum, automorphic equivalence represents the limiting case of exact invariance under all vertex permutations that preserve adjacency and weights, whereas structural and regular equivalence more naturally admit graded, approximate similarity. Comparing direct connections or comparing roles of neighbours easily extends into quantitative measures, but a notion of continuous automorphic distance was also proposed for unweighted networks~\cite{automorph_distance}. Applying automorphic equivalence to weighted and noisy networks is less explored. 

Exact automorphisms are fragile: even minor perturbations can destroy them, particularly in weighted or directed graphs. Conversely, approximate or embedding-based methods often blur algebraic symmetries into statistical similarity, losing interpretability. Bridging this gap requires frameworks that retain the rigor of automorphism-based analysis while accommodating the continuous and uncertain nature of empirical data. Our method contributes to this effort by enabling symmetry analysis through controlled weight coarsening, preserving structural invariance at interpretable levels of approximation.

A variety of frameworks have explored approximate symmetries in another sense, by tolerating missing or extra edges. These studies range from three-dimensional geometry~\cite{Mitra_2006} and differential equations~\cite{Pakdemirli_2004} to networks with nearly symmetric adjacency patterns~\cite{Liu_2020, Pidnebesna_2025, Rosell-Tarrago_2021}. They quantify how closely a system can be permuted onto itself, often using optimization or dynamical models to measure deviation from exact invariance. However, in e.g. biological weighted networks, even few edges with small weights can have functional significance. In ecological systems weak links can play a crucial stabilizing role~\cite{McCann1998, Jacquet2016}, so their contribution cannot be ignored.

To address this, our framework applies the principle of network symmetry to weighted graphs through \textit{weight aggregation} and does not neglect any edge. Edge weights are iteratively approximated by an aggregation function chosen according to the observed edge-weight distribution. To retain interpretability and consistency across levels of approximation, the resulting edge groups are constructed so that each coarser partition contains the finer ones. If two edges belong to one category at some aggregation level, they also belong to one category as the discretization proceeds. This produces a nested sequence of increasingly simplified versions of the original network, on which exact symmetries can be studied. The aggregation function itself is arbitrary and can be adapted to the data at hand--for example, based on order of magnitude, quantiles, or uniform binning--allowing flexibility across empirical systems. Conceptually, this procedure parallels \textit{conceptual scaling} in Formal Concept Analysis (FCA)~\cite{Ganter_Wille}, where continuous attributes are discretised to reveal latent structural patterns.

Symmetries in networks also offer computational advantages by reducing redundancy: computations performed on one vertex can generalise across all vertices in its orbit. This principle has been used in unweighted networks to simplify large-scale analyses, reduce matrix dimensionality, and optimise algorithmic performance~\cite{HolmeDetectingDegree, MacArthur20083525, Xiao2008}. Observed symmetries can even explain emergent structural patterns, such as correlations between degree and connectivity targets~\cite{MacArthur20083525, Xiao2008}. Beyond methodological benefits, automorphic symmetries have been used to identify invariant substructures in unweighted physical~\cite{SanchezGarcia2020} and biological systems, brain networks~\cite{Hu2014}, and social graphs.

In ecology, network symmetries can reveal \textit{functionally substitutable species}, organisms that occupy equivalent trophic positions or interact with the ecosystem in analogous ways. Physically, species and whole ecosystems rely on flows of organic matter (biomass) that sustain them. As they constitute the most basic layer of ecological relationships, food webs were the primary tools of ecological similarity studies. A weighted food web is a directed graph presenting the flows of biomass between functional groups of species, non-living organic matter pools within an ecosystem and their exchange with the external environment.

This approach complements traditional pairwise similarity measures, such as trophic niche overlap indices~\cite{Pianka_niche_overlap}, that estimates the similarity of predators and prey and implements structural equivalence in ecology. The concept of substitutability in food webs has deep ecological implications: it relates to coexistence mechanisms~\cite{Gause1934, Chesson_2000, Chesson_2018, Levine_2017} and connects to classical notions of the realised niche~\cite{Hutchinson, Macarthur_realised_niche_1967}.

This article bridges methodological advances in graph theory with ecological theory. In food webs, network symmetries complement established aggregation techniques such as trophic species grouping~\cite{Yodzis1999, Williams2000}, functional group modeling~\cite{Fulton2003, Brose2006}, and community detection~\cite{Newman2006, Fortunato2010}. All of these approaches aim to reduce complexity by identifying sets of vertices that play similar functional roles. Automorphism-based analysis detects these equivalences directly from network topology, providing a mathematically grounded lens on functional redundancy and substitutability in ecosystems.

In Section~\ref{sec:methods}, we introduce our weight approximation framework, food webs, and our choice of aggregation function adapted to their empirical edge weight distribution. We introduce graph measures that use symmetric vertices, orbits and automorphism group structure to quantify how symmetric a network is, and trophic level, a metric aiding the interpretation of orbit properties in ecological context. In Sec.~\ref{sec:results} we analyse the proposed method both in a case study, applied to one real-world network, and through properties of orbits and values of symmetry measures observed in the whole dataset. We discuss our method in Sec.~\ref{sec:discussion}, and conclude with Sec.~\ref{sec:conclusions}.

\section{Methods and data}\label{sec:methods}
This section defines the key graph theoretical concepts, data, and tools used in our analysis. We introduce automorphisms, orbits, and the approximation of energy flows into weight classes, along with the symmetry measures applied. We introduce food webs and briefly outline the computational methods used.

\subsection{Automorphisms and orbits}
For a directed graph $G=(V,E)$ with a real-valued weight function $\phi: E \to \mathbb{R}$, a graph automorphism $\sigma: V \to V$ is a permutation of the vertices such that $\phi_{(u,v)} = \phi_{(\sigma(u),\sigma(v))}$ for all $u, v \in V$\footnote{For visual clarity, we put the argument of the weight function in the lower index.}. 

The set of all automorphisms $\rm{Aut}(G)$ of a graph $G$ forms a group. Its action on the vertex set $V$ always partitions it into disjoint equivalence classes called orbits. An orbit of a vertex $v \in V$, $\rm{Aut}(G)\cdot v=\{a\cdot v: \hspace{0.2cm} a\in \rm{Aut}G\}$ corresponds to a set of vertices that can be permuted without altering the graph structure. An orbit has only one element if no automorphisms map this element to any other vertex. 

An orbit containing more than one vertex will be called a \textit{symmetric orbit} and its elements \textit{symmetric vertices}. Formally, we say that two vertices $v,w$ are symmetric if there exists an authomorphism $\sigma\in Aut(G)$ such that $\sigma(v)=w$. We denote the number of orbits as $N_O$. The number of symmetric vertices $N_S$ equals also the number of elements in orbits containing more than one element.

\subsection{Aggregating flows into classes}
\label{subs: aggregation}
In empirical weighted networks, symmetry might be hidden by the fact that two independently measured edge weights are typically distinct real numbers. If taken at face value, such uniqueness would imply that no two vertices are exactly symmetric: each edge weight difference breaks potential automorphisms. Existing exact symmetries might also reflect measurement or methodology artifacts or discrete modeling choices. 

To robustly study regularities in weighted systems, we introduce a family of weight aggregation functions
$\Phi_\alpha: \phi[E] \rightarrow \mathbb{Z}$ parameterised by a coarseness level $\alpha \in \mathbb{N}$, which map continuous weights into discrete categories. Each aggregation $\Phi_\alpha$ defines an approximated network whose automorphism group captures approximate symmetries of the original weighted network. For brevity, we shall also denote aggregated versions of the original network under consideration and their weight functions just by $\Phi_\alpha$. 

The family $\Phi_{\alpha}$ provides a bridge between two extremes:
\begin{itemize}
    \item for $\alpha \to 0$, the discretization approaches the fully weighted case, recovering the original network,
    \item for sufficiently large $\alpha$, all weights collapse to a single category, yielding the unweighted version $\Phi_{\infty}$ of the original network, and its corresponding symmetries.
\end{itemize}

We require that the categories are nested, i.e.
\begin{equation}
(\Phi_\alpha(e) = \Phi_\alpha(e')) \Rightarrow (\forall_{\alpha'>\alpha} \Phi_{\alpha'}(e) = \Phi_{\alpha'}(e') ).
\end{equation}
This also means, that if $a\in \rm{Aut}(\Phi_\alpha)$, then $a\in \rm{Aut}(\Phi_{\alpha'})$ for all $\alpha'>\alpha$. In other words, the automorphism group at some aggregation level is a subgroup of the automorphism groups at coarser aggregations, including the fully unweighted version $\Phi_\infty$, that does not depend on the choice of a particular aggregation function.
 
Traversing this family therefore traces a path from the exact orbits of the original weighted network to those of its unweighted counterpart. The aggregation parameter $\alpha$ plays a role analogous to a resolution parameter, controlling the scale at which structural similarity is detected. While the quantities such as the level $\alpha$ at which orbits appear, the order of appearance and merging together depend on the chosen aggregation function, the endpoints of the path are fixed. The evolution of orbits along this path offers insight into how vertex similarity emerges as weight distinctions are progressively relaxed.

The differences between such orbit paths traced by different aggregation functions are typically small due to the constraint given by the symmetries of the fully unweighted network, but unavoidably remain as a direct consequence of binning. Exploring several complementary aggregation functions may shed light on the subgroups of $\Phi_\infty$ automorphisms realised in some approximations of the studied network.

\subsection{Logarithmic aggregation}
\label{subs: log_agg}

Biomass flows in food webs vary across several orders of magnitude and follow an approximately log-normal distribution~\cite{Bascompte2005, Albatross}, therefore logarithmic aggregation offers a natural and scale-consistent discretisation of edge weights. By grouping together interactions that differ by less than one or more orders of magnitude, we construct a sequence of approximations $\Phi_\alpha$ ($\alpha = 1, 2, 4, \dots$) that reflect ecological uncertainty and measurement variability while revealing robust structural symmetries.  
\be\label{eq:aggregation}
    \Phi_\alpha (\phi_{e}) = 
    \begin{cases}
    \phi_{e} \; &\text{for} \;\; \alpha = 0,\\
    1+\lfloor\frac{ log_{10}(\phi_{max})-log_{10}(\phi_{e})}{\alpha}\rfloor \;\; &\text{for} \;\; \alpha = 1,2,4, ... \\
    1 \; \; &\text{for} \; \; \alpha = \infty.
    \end{cases}
\ee
where $\phi_{e}$ is the weight of an edge $e\in E$ and $\phi_{max}$ is the largest weight in the network. $\Phi_0$ is an identity function, and $\Phi_\infty$ assigns the same weight to each edge, converting a graph to its unweighted simplification. To ensure that flows grouped together by a previous aggregation belong to the same set in the following ones, we define $\alpha$ values as powers of two. In practice $\Phi_8\sim\Phi_\infty$, as only few food webs contain flows that differ by more than eight orders of magnitude.

This construction ensures that vertices within an orbit of $\Phi_{1}$ differ in edge weights by no more than one order of magnitude, within an orbit of $\Phi_{2}$ by two, and so forth. In what follows, we explore the structural patterns revealed by such progressively coarsened approximations, noting that the resulting symmetries represent consistent and interpretable approximations of the weighted system. 

The aggregation defined in Eq.~\ref{eq:aggregation} is not adaptive to the empirical weight distribution of a given network, but instead imposes a fixed partition of the weight space anchored at $\phi_{max}$. This reflects a deliberate choice: comparing multiple networks requires a scheme that is consistent across datasets rather than tailored to each distribution. The same aggregation scheme is also applied irrespective of whether the underlying distribution is unimodal, multimodal, or uniform on the logarithmic scale. The aggregation therefore defines a parametrised family of scale transformations, capturing robustness of orbit structure under controlled coarsening of interaction magnitudes, rather than distribution-specific features.

In a case study, we further compare alternative normalizations of this aggregation, $\Phi_{\alpha}^{\text{min}}$ changing the normalisation factor from the maximal weight $\phi_{max}$ in $\Phi_{\alpha}$~\ref{eq:aggregation} to the minimal $\phi_{min}$, and $\Phi_{\alpha}^{\text{med}}$ using the median weight $\phi_{med}$. This way, we assess how normalisation can affect the sequence in which vertex orbits emerge along the path from the original weighted network to its unweighted form.

\begin{figure}[H]
    \centering
    \includegraphics[width=0.95\textwidth]{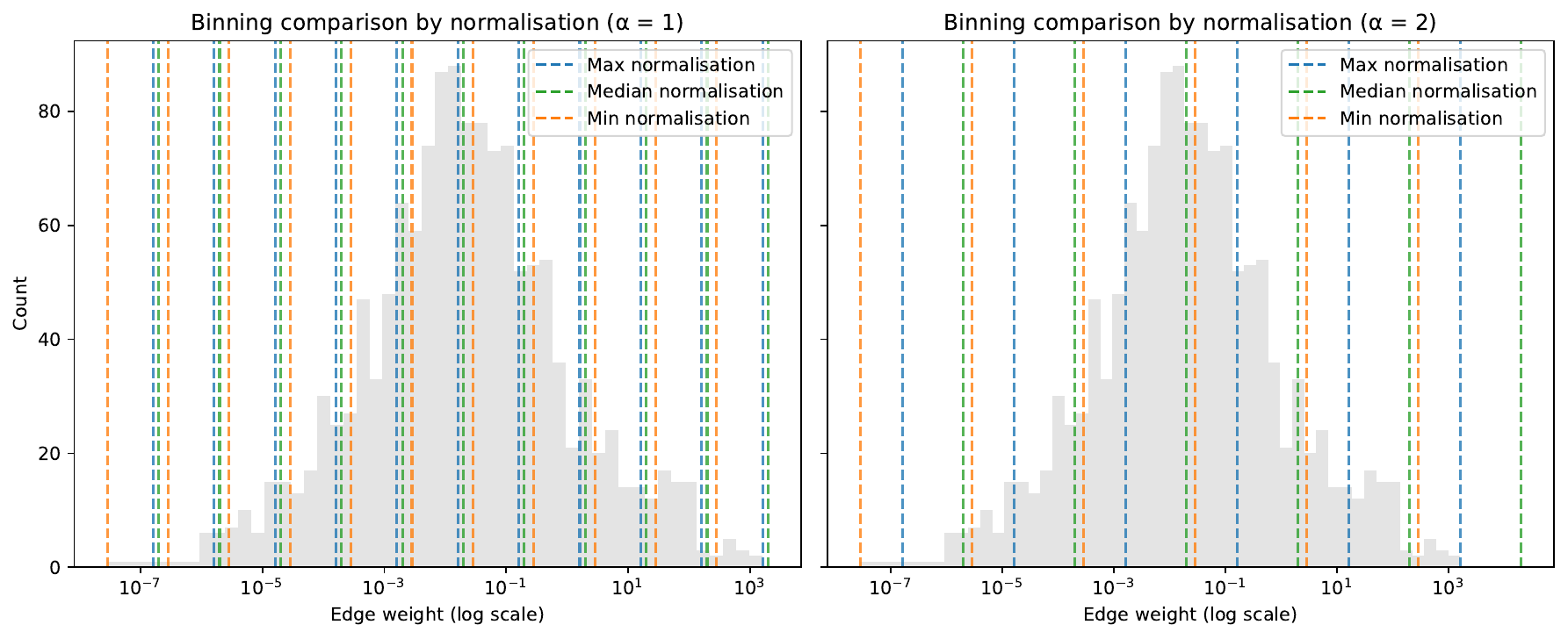}
    \caption{A log-scale histogram of edge weights in the Albatross Bay food web~\cite{Albatross}. Vertical lines mark divisions between logarithmic bins in the first two aggregations: blue for $\Phi_{\alpha}^{\text{max}}$, green for $\Phi_{\alpha}^{\text{med}}$, orange for $\Phi_{\alpha}^{\text{min}}$, with  $\alpha=1$ on the left and $\alpha=2$ on the right.}
    \label{fig:hist_weights_bins}
\end{figure}

We illustrate the concept of alternative normalisations for one of the largest analysed food webs~\cite{Albatross} in Fig.~\ref{fig:hist_weights_bins}, showing edge weights and bins in the three normalisations for the first two weight aggregations.

\subsection{Symmetry measures}

While the analysis of orbits and symmetric vertices reveals properties of individual vertices within a network, \textbf{graph symmetry measures} quantify the overall symmetry of the entire graph. This allows meaningful comparisons between different networks.

Let us consider a graph $G$ with $N$ vertices, $N_O$ orbits, and $N_S$ symmetric vertices.

\begin{figure}[H]
    \centering
    \includegraphics[width=0.95\textwidth]{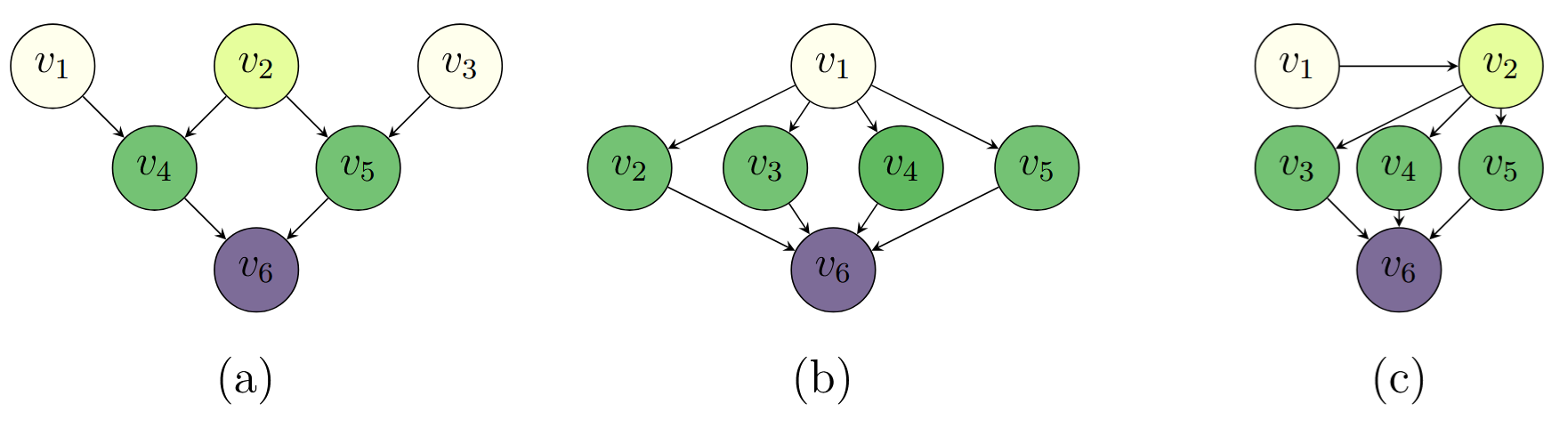}
    \caption{Three example directed graphs with the same number of vertices but with a different structure of symmetry. Different colours correspond to different automorphism group orbits within each graph. Orbits of (a): $\{v_1, v_3\}, \{v_4, v_5\}, \{v_2\}, \{v_6\}$  (b): $\{v_2,v_3,v_4,v_5\}, \{v_1\}, \{v_6\}$, (c): $\{v_3,v_4,v_5\}, \{v_1\}, \{v_2\}, \{v_6\}$. One-element orbits indicate lack of any symmetry transforming the vertex they contain. Graphs (a) and (b) have the same number of symmetric vertices (4) but a different number of orbits (4 vs 3).  On the other hand, graphs (a) and (c) have the same number of orbits, but a different number of symmetric vertices (4 vs 3).}
    \label{fig:example_graphs}
\end{figure}

\textbf{The symmetric vertices ratio}, introduced in \cite{Xiao2008}, is defined as:
\begin{equation}
    S_V = \frac{N_S}{N}
\end{equation}
This measure describes the proportion of vertices that are symmetric with respect to some nontrivial graph automorphism. Despite its intuitive appeal, graphs with differing symmetry structure can have the same value of $S_V$.

Consider the graphs shown in Fig.~\ref{fig:example_graphs}, each with $N=6$. Graphs (a) and (b) have $S_{V,a} = S_{V,b} = \frac{4}{6} = 0.67$, while $S_{V,c} = \frac{3}{6} = 0.5$. Yet, $S_{V}$ does not describe whether the symmetric vertices are connected through multiple independent automorphisms or just a few complex ones.
\textbf{Redundancy}, proposed in~\cite{MacArthur20083525}, takes a complementary approach by focusing on the number of orbits:
\begin{equation}
r = \frac{N_O - 1}{N}, \quad r \in [0, 1)
\end{equation}
This measure reaches its maximum when every vertex belongs to its own orbit (no symmetry), and is zero when all vertices are fully symmetric (i.e., belong to a single orbit). We note it runs against the general notion of redundancy, but follow the convention proposed by measure's proponents.

Redundancy can yield different interpretations than $S_V$. For example, while graph (c) has a lower $S_V$ than graph (a), they both have the same redundancy: $r_1 = r_3 = \frac{4 - 1}{6} = 0.5$. Graph (b) — previously tied with graph (a) in $S_V$ — has lower redundancy: $r_2 = \frac{3 - 1}{6} = 0.33$, suggesting it is more symmetric under this measure.

\textbf{The Beta measure}, proposed in~\cite{MacArthurAndersonBeta}, is defined as:
\begin{equation}
\beta = \left(\frac{|\text{Aut}(G)|}{N!}\right)^{1/N}
\end{equation}
Here, $|\text{Aut}(G)|$ is the size of the graph’s automorphism group, and $N!$ is the total number of possible permutations of $N$ vertices. Thus, $\beta = 1$ corresponds to a fully symmetric graph, while $\beta = 0$ indicates a completely asymmetric one.

This measure is sensitive to the structure of the graph. In graph (b), vertices $v_1$ and $v_6$ are asymmetric, but vertices $v_2$ through $v_5$ can be freely permuted, yielding $4!$ automorphisms. Hence,

\[
\beta_2 = \left(\frac{1! \cdot 1! \cdot 4!}{6!}\right)^{1/6} \approx 0.567.
\]
Similarly, graph (c) has $3!$ automorphisms, resulting in $\beta_3 \approx 0.45$.

However, the symmetries in graph (a) are more constrained. For example, if $v_1 \mapsto v_3$, symmetry requires $v_4 \mapsto v_5$. But if $v_1$ maps to itself, then $v_4$ must also map to itself. Since $v_2$ and $v_6$ are asymmetric, only two automorphisms are allowed: the identity and the one swapping $(v_1, v_3)$ and $(v_4, v_5)$. Thus, $\beta_1 = \left( \frac{2}{6!} \right)^{1/6} \approx 0.375$.

These three measures capture basic properties of graph symmetry. In this article we explore their usefulness in real-world examples that in contrast to our simple examples in Fig.~\ref{fig:example_graphs} differ in vertex number $N$. While $\beta$ distinguishes graphs of the same size effectively, we shall see that it becomes less informative when $N$ varies substantially.

\subsection{Food webs}

A food web represents an ecosystem as a weighted and directed graph with species or functional groups of species mapped to vertices connected by flows of biomass resulting from feeding relationships. The most common models describe mass flows (expressed in dry or wet weight), or closely related flows of carbon. Only a few models quantify the flow of nitrogen or phosphorus.

In this analysis we restrict ourselves only to local interactions between vertices representing living or non-living pools of biomass (detritus), ignoring exchanges with outside environment, such as atmosphere. These are usually also included in food web models as the import, export and respiration of each vertex.

\subsection{Trophic level} 

Within this network representation, trophic level is one of the earliest and most widely used global descriptors of vertex position. It was introduced as an ordering variable in food chains and ecological pyramids~\cite{Elton}. It places vertices along the dominant direction of biomass flow, counting feeding steps from the primary sources of biomass to a given vertex and further to top predators. 

It was formalised further by~\cite{Lindeman}, showing that energy transfer tends to diminish systematically with each successive feeding step. Parallel conceptual developments--notably the formalisation of the ecological niche~\cite{Hutchinson}--emphasised interconnectedness of ecological processes and the range of conditions under which species persist or interact.

Subsequent work on ecological complexity, including on food web topology and stability, has shown that strictly discrete feeding categories are often inadequate to describe real systems. It motivated continuous measures of trophic position~\cite{Pimm1982} that more faithfully capture network features such as omnivory and cycles.

In such settings, assigning a trophic level to each vertex is inherently non-trivial, particularly in cyclic networks. While early approaches relied on discrete classifications or shortest-path distances from basal resources, modern formulations~\cite{Pauly1998} define trophic position recursively, allowing non-integer values that more faithfully capture the structure of real food webs.

In this paper we follow \cite{Pauly1998}, and assign trophic level 1 to non-living components and primary producers. For all other vertices, we define the trophic level recursively as a biomass-weighted average of the levels of their resource vertices. If $\phi_{ji}$ denotes the biomass flow from vertex $j$ to vertex $i$, then the trophic level $\tau_i$ is given by
\begin{equation}\label{trophic_l}
\tau_i := 1 + \sum_{j=1}^N \frac{\phi_{ij}}{\sum_{k=1}^N \phi_{ki}} \tau_j.
\end{equation}

For a connected food web, a unique solution exists if 1 is not an eigenvalue of the diet proportion matrix $\Bigl (\frac{\phi_{ij}}{\sum_{k=1}^N \phi_{ki}} \Bigr )_{i,j}\in \mathbb{R}^{N \times N}$.

\subsection{Dataset and automorphism computation}
The analysed food webs, primarily describing marine ecosystems, originate from models published by ecosystem research groups. We provide their full bibliography in Supplementary Material. Most datasets were compiled from Ecopath with Ecosim~\cite{EwE} and retrieved from Ecobase~\cite{Ecobase}.

To calculate food web automorphisms and their properties, we use the open source software SageMath~\cite{sage}.

\section{Results}\label{sec:results}
We apply the weight approximation method, see Eq.~\ref{eq:aggregation}, to explore symmetries in a dataset of 250 empirical food webs. While the raw data reveals almost no symmetries, many symmetric structures appear once flow values are approximated. We compare these results across different stages of aggregation to see how symmetry depends on the modelling approach.

Since each network vertex corresponds to a real species within an ecosystem, we focus on the characteristics of symmetric vertices and their orbits. We analyse the size and structure of these orbits and consider how they relate to ecological roles reflected by trophic levels. Finally, we compare three symmetry measures to evaluate how much distinct information they capture.

\subsection{Case study: Peruvian upwelling food web}

\begin{figure}[H]
    \centering
    \includegraphics[width=0.95\textwidth]{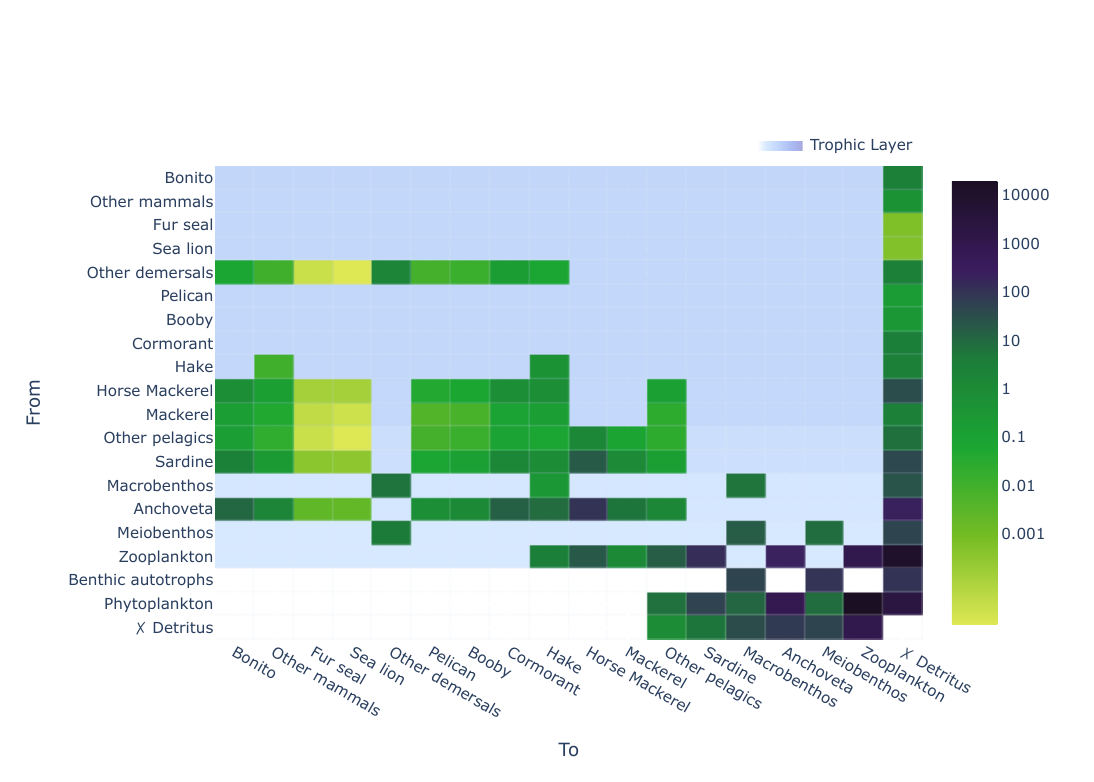}
    \caption{A heatmap representing the adjacency matrix of the Peruvian upwelling food web, drawn with foodwebviz~\cite{foodwebviz}. The color of a cell maps the biomass flow from the vertex in the row to the vertex in the column. Vertices are sorted according to their trophic level, indicated by the shades of blue of the table background.}
    \label{fig: Peru_heatmap}
\end{figure}

\begin{figure}[H]
    \centering
    \includegraphics[width=0.95\textwidth]{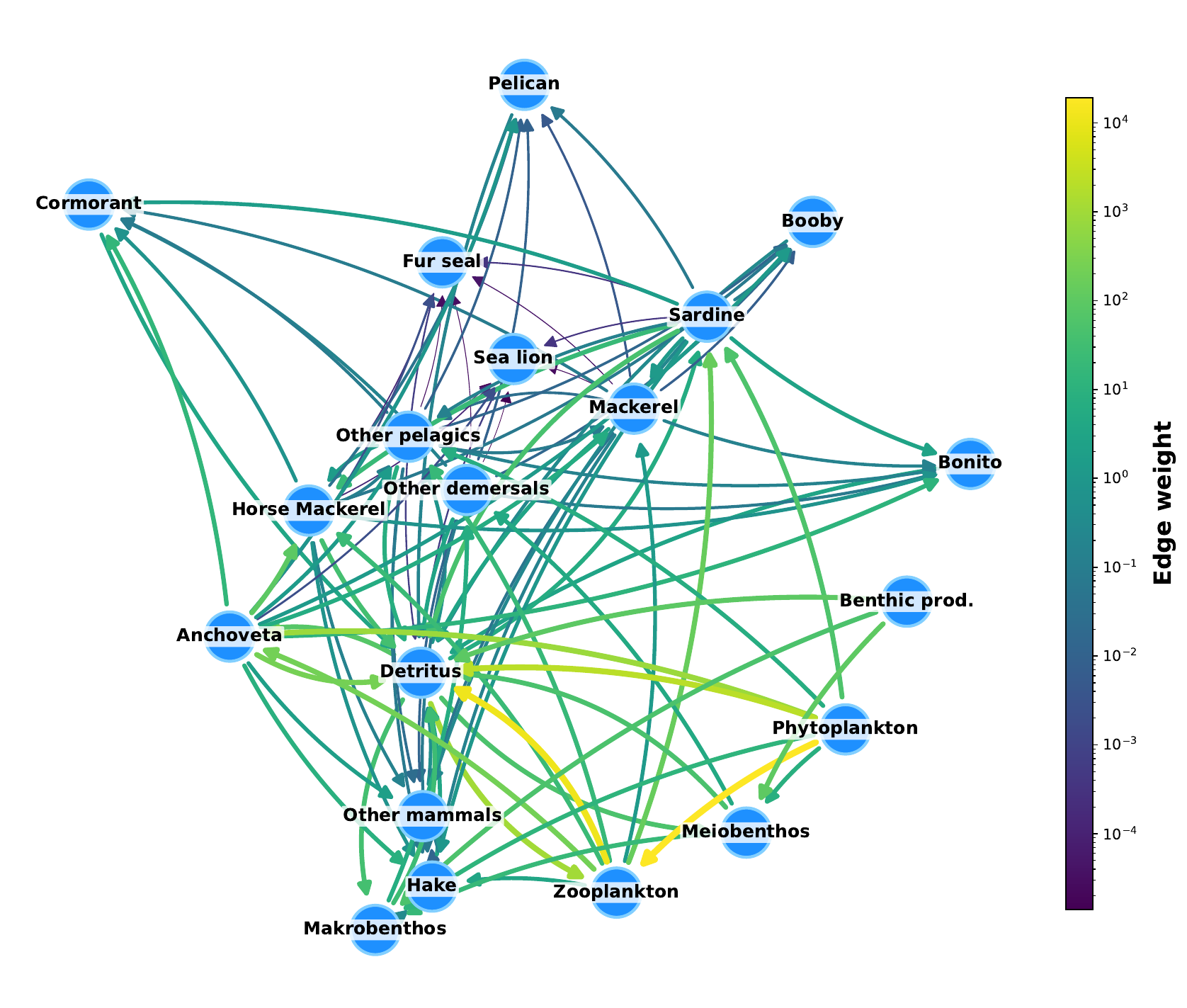}
    \caption{Peru upwelling food web~\cite{Peru} drawn in the graph form. Colours and edge widths use a logarithmic map to represent edge weights.}
    \label{fig: Peru_diagram}
\end{figure}

An analysis of automorphism orbits in a specific ecological network can reveal structural patterns and functional roles. We demonstrate this using the Peruvian upwelling food web~\cite{Peru}, with Ecobase ID 311 (see Fig.~\ref{fig: Peru_heatmap} for its adjacency matrix presented as a hetmap, Fig.~\ref{fig: Peru_diagram} for its traditional diagram, and Supplementary Material for its animation).

The Peruvian upwelling food web consists of 20 vertices, making it relatively small compared to other food webs in the dataset. It relative simplicity allows observing the symmetric vertices directly in the structure. Without aggregation and under aggregation $\Phi_1$, this food web does not contain any symmetric orbits. Disregarding weight differences smaller than two and four orders of magnitude (in $\Phi_2$ and $\Phi_4$), two pairs of vertices are symmetric: $O_1^{\Phi_2} = \{$Booby, Pelican$\}$ and $O_2^{\Phi_2} = \{$Fur seal, Sea lion$\}$. Table~\ref{tbl: PeruOrbits} presents their mean trophic level and degree. The first orbit consists of two piscivorous bird species, while the second groups two species of pinnipeds. The unweighted graph $\Phi_{\infty}$ contains a large orbit that completely merges the widened group of piscivorous birds with mammals, but also predatory bonito fish: $O_1^{\Phi_{\infty}} = \{$Fur seal, Sea lion, Booby, Pelican, Bonito, Cormorant$\}$. Another orbit connects two mackerel species $O_2^{\Phi_{\infty}} = \{$Mackerel, Horse Mackerel$\}$. A third orbit $O_3^{\Phi_{\infty}} = \{$Sardine, Anchoveta$\}$ signals a functional similarity between two planktivorous fish, of which sardines are generally more omnivorous, while anchoveta sustain much higher biomass flows.

\begin{table}[h]
\centering
\scalebox{0.75}{
\renewcommand{\arraystretch}{1.3}
\begin{tabular}{l l l | l l l} 
    \multicolumn{3}{c!{\color{gray}\vrule}}{$\Phi_2$} &  \multicolumn{3}{c}{$\Phi_{\infty}$} \\
    Orbits & Mean t.l. & $\deg(v)$ & Orbits & Mean t.l. & $\deg(v)$ \\
    \hline
     \{Fur seal, Sea lion\} & 3.35 & 7  & \{Fur seal, Sea lion, Bonito, & &  \\ 
     \{Booby, Pelican\} & 3.32 & 7 & Booby, Pelican, Cormorant\} & 3.34 & 7 \\
    - & - & - & \{Mackerel, Horse Mackerel\} & 3.26 & 14 \\
    - & - & - & \{Sardine, Anchoveta\} & 2.47 & 15 \\
\end{tabular}
}
\caption{Symmetric orbits present in Peruvian upwelling food web under different aggregations. Columns \textit{Mean t.l.} contain mean trophic level of vertices that are elements of orbits $O_1,O_2,O_3$. We omit columns $\Phi_0, \; \Phi_1$ as the Peruvian upwelling food web is asymmetric under those aggregations, and $\Phi_4$ as it contains exactly the same symmetries as $\Phi_2$.}
\label{tbl: PeruOrbits}

\end{table}

\begin{figure}[H]
    \centering
    \subfloat[]{
	\includegraphics[width=0.48\linewidth]{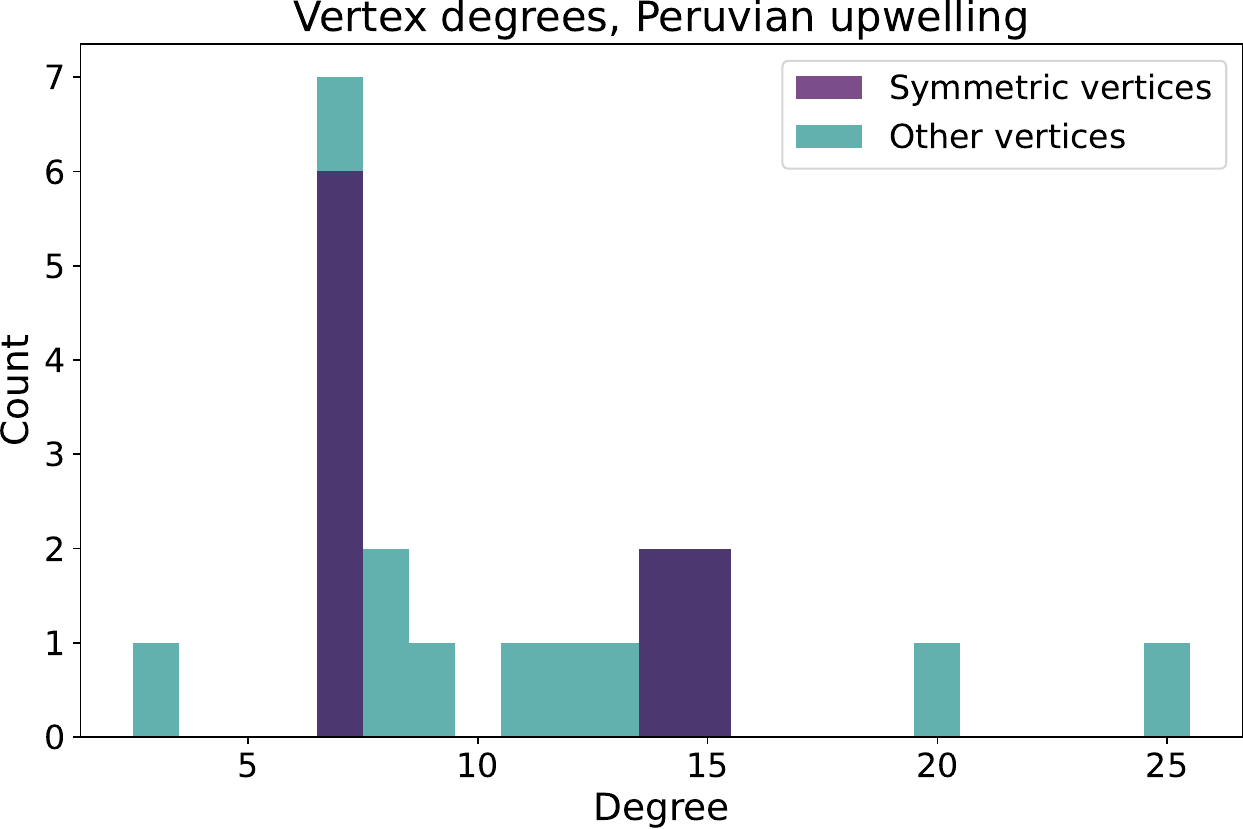}}
    \subfloat[]{
	\includegraphics[width=0.485\linewidth]{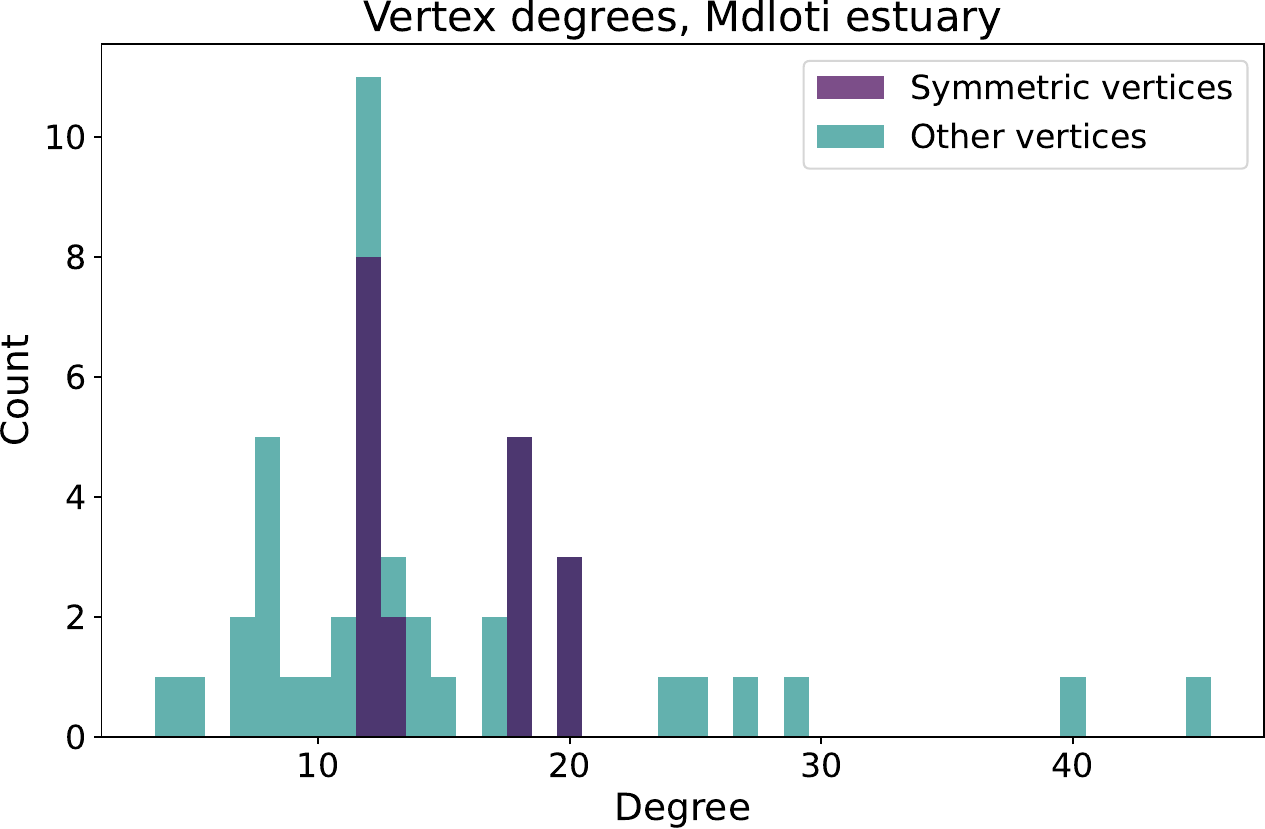}}
    \caption{Histograms of vertex degrees in food webs of Peruvian upwelling (left), and Mdloti estuary (right). The contributions of symmetric vertices are violet; of other vertices--turquoise.}
   
    \label{fig:degree_hists}
\end{figure}

All symmetric vertices in this case have a total degree of seven or more. This is somewhat surprising, as one might expect that vertices with many connections would be harder to match and less likely to be part of a symmetric orbit. This has two main reasons, the highly nonrandom structure of food webs, and approximation made by the modeller, who might have neglected less important edges. The Peruvian Upwelling model might indeed be such a rough approximation, but the same pattern of highly connected vertices can be found in more detailed food webs, e.g. of Mdloti estuary~\cite{Mdloti}. The degree distributions highlighting the contributions from symmetric vertices are shown in Fig.~\ref{fig:degree_hists}.

There are important hidden variables, such as the physical size of considered organisms. Two species of similar size feed on species of sizes that are also similar. Physical and biological constraints limit the possibility of extending their interactions to completely different organisms. In this way, orbits may detect functional aspects of food webs that are shaped by such variables, even though these variables are not explicitly represented in the food web.

Table~\ref{tbl: PeruMeasures} presents the values of symmetry measures applied to the weight approximated copies of the Peruvian upwelling food web. As we will see later in the text, the values of redundancy and the beta measure for the Peruvian upwelling food web under aggregations $\Phi_2$ and $\Phi_4$ are typical for its size (see Fig.~\ref{fig: hist_measures_vertices}). However, under aggregation $\Phi_{\infty}$, $50\%$ of all vertices are considered symmetric. Peru's redundancy of $0.6$ is among the lowest in the dataset.

\begin{table}[h]
    \centering
    \scalebox{1}{
\begin{tabular}{l |c c c !{\color{gray}\vrule}  c c c !{\color{gray}\vrule} c c c !{\color{gray}\vrule} c c c}
    & \multicolumn{3}{c!{\color{gray}\vrule}}{ $\Phi_2$} & \multicolumn{3}{c!{\color{gray}\vrule}}{$\Phi_4$} & \multicolumn{3}{c!{\color{gray}\vrule}}{$\Phi_{\infty}$} \\
    & $r$ & $S_V$ & $\beta$ & $r$ & $S_V$ & $\beta$ & $r$ & $S_V$ & $\beta$ \\
    \hline
    Peru  & 0.85 & 0.2 & 0.13  & 0.85 & 0.2 & 0.13 & 0.6 & 0.5 & 0.18 \\
\end{tabular}
}

    \caption{Values of symmetric measures symmetric vertices ratio $S_V$, redundancy $r$ and the beta measure $\beta$ of Peruvian upwelling food web under different aggregations.}
    \label{tbl: PeruMeasures}
\end{table}

We compare non-trivial orbits on the path of aggregation function family $\Phi_{\alpha}$ with other normalisation choices - to the minimal $\Phi_{\alpha}^{\text{min}}$ and median $\Phi_{\alpha}^{\text{med}}$ flow instead of the maximum. This effectively shifts the discretisation bins without changing their size. $\Phi_{\alpha}^{\text{min}}$ exhibits exactly the same orbits at various aggregations $\alpha$ as $\Phi_{\alpha}$. $\Phi_{2}^{\text{med}}$ and $\Phi_{4}^{\text{med}}$ also contain $O_2^{\Phi_2} = \{\text{Fur seal, Sea lion}\}$, but replace the $O_1^{\Phi_2} = \{\text{Booby, Pelican}\}$ orbit with $\{\text{Bonito, Cormorant}\}$. This shows that also $\{\text{Bonito, Cormorant}\}$ symmetry depends on flows that differ by at most two orders of magnitude. Both pairs are parts of $O_1^{\Phi_{\infty}}$ as orbits of the unweighted version of the network are independent of any aggregation procedure and contain candidates for more accurately symmetric vertices. The number of symmetric vertices, number of orbits and their sizes do not change, therefore these different normalisations have no impact on any graph symmetry measure in this case. 

In fact, any discretisation method of continuous intervals retains such ambiguity. Values close to the boundary of discrete categories but on its opposite sides will be separated despite being similar. The aggregation method is a simple way to find true similarities existing up to a given approximation also in the original network with existing graph automorphism algorithms, but is not exhaustive. Different aggregation functions or their variants might uncover other approximate symmetries that are subgroups of the symmetry group of the unweighted network $\rm{Aut}(\Phi_{\infty})$. Each approximate symmetry is tied to the aggregation function used and carries associated interpretation. The order-of-magnitude aggregation function is tied to the physical values and gives a universal interpretation. Discretisation methods based e.g. on quantiles of the observed distribution depend on weight distribution of a particular network. Their interpretation will thus differ between various networks.

\subsection{Case study: Mdloti estuary food web}
\begin{figure}[H]
    \centering
    \includegraphics[width=0.95\textwidth]{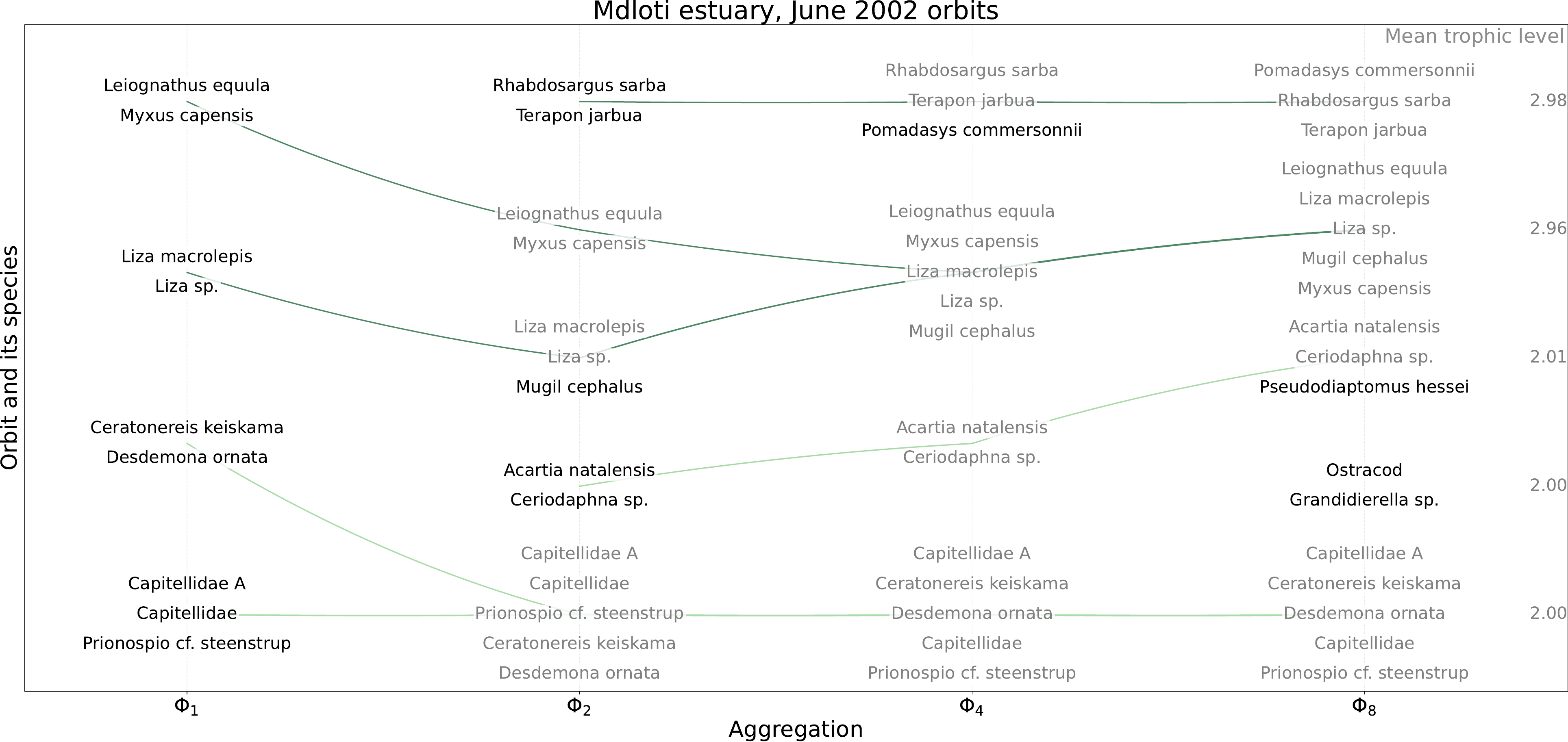}
    \caption{Species in orbits of Mdloti estuary food web approximations. The first appearance of a species is black, and later in aggregation progress is grey. Orbits are ordered vertically by the mean trophic level of their vertices.}
    \label{fig:mdloti_orbits}
\end{figure}

While the Peruvian upwelling food web was small enough to discuss individual species in orbits, the medium-sized food web of Mdloti estuary in South Africa sampled in June 2002~\cite{Mdloti} offers a slightly more elaborate illustration of how orbits appear and merge at various aggregation levels. $\Phi_{0}$ has no symmetric vertices. Fig.~\ref{fig:mdloti_orbits} presents species appearing in orbits of $\Phi_{\alpha}^{\text{max}}$, $\alpha \in \{1,2,4,8\}$.

Two $\Phi_{8}$ orbits at trophic levels close to 3 contain fish species, two others--tiny crustaceans, and the bottom one--polychaetes (bristle worms). We see how the gradually relaxed weight differences lead to merging of orbits. Species belonging to the same orbit of $\Phi_{1}$ can be expected to be more similar than ones that come together only later.

\subsection{Properties of orbits}

In this section, we analyse the set of all orbits in the consecutive aggregations of the 250 empirical food webs.
We observe that the order of the automorphism group equals the product of factorials of its vertex orbit sizes--consistent with the group decomposing as a direct product of symmetric groups, each acting on one orbit. This means every permutation within an orbit extends to a valid automorphism while leaving all other orbits fixed. Consequently, automorphic equivalence coincides with structural equivalence in the studied food webs.

\subsubsection{Size}
\begin{figure}[H]
    \centering
    \subfloat[]{
	\includegraphics[width=0.48\linewidth]{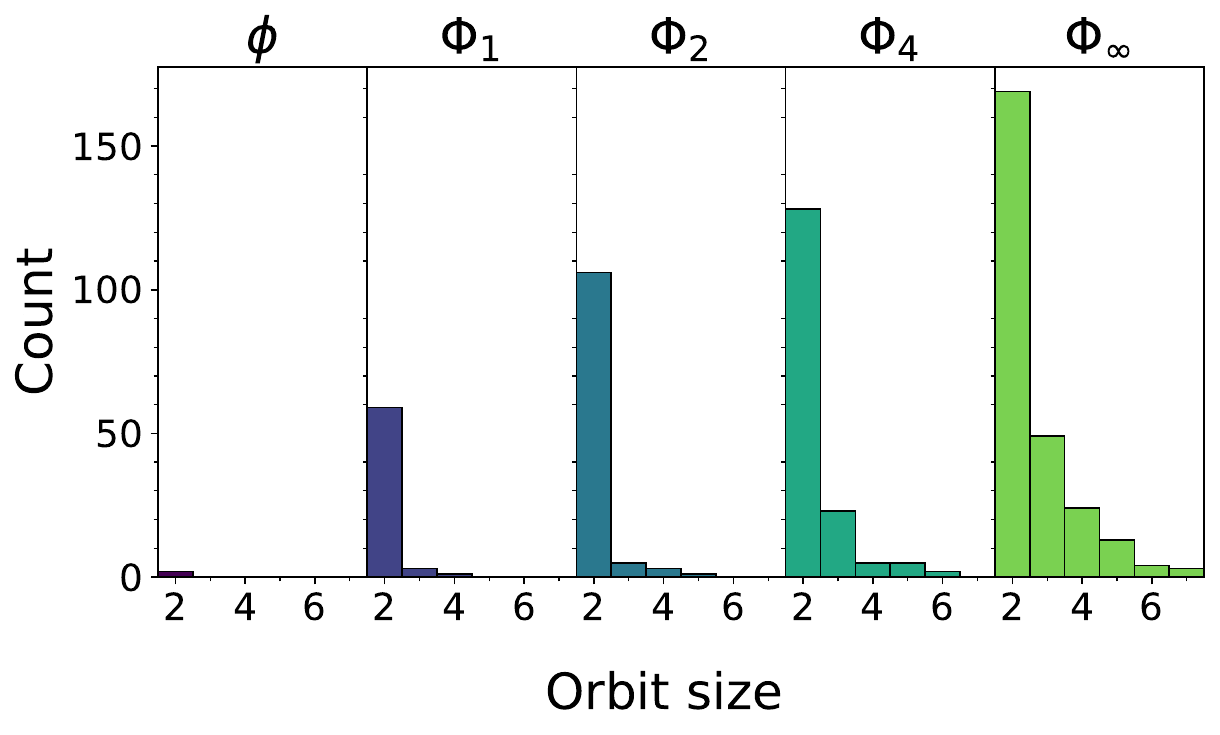}}
    \subfloat[]{
	\includegraphics[width=0.52\linewidth]{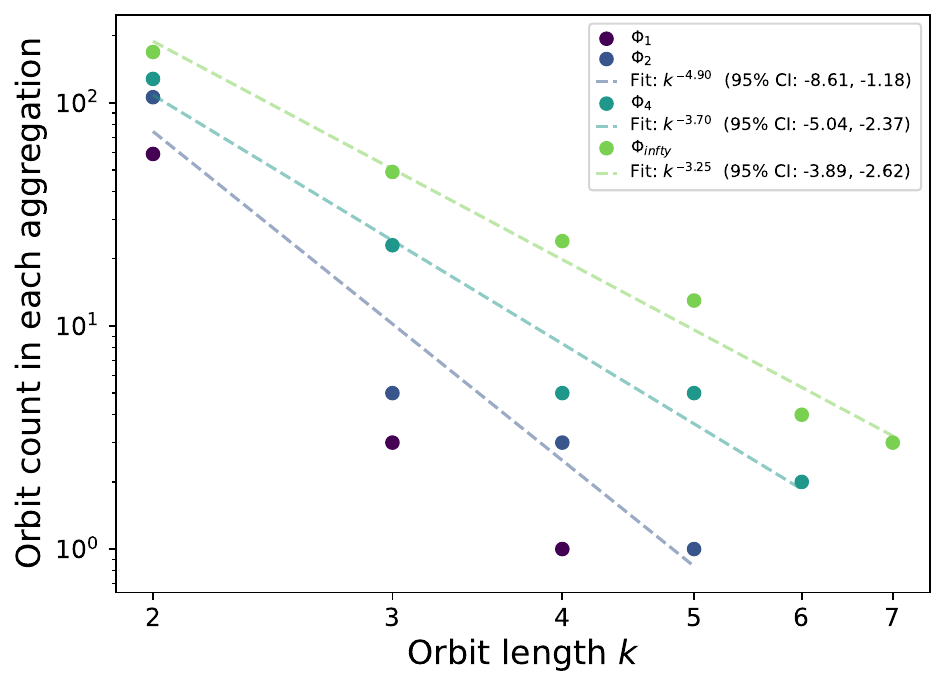}}
    \caption{Left: a histogram of symmetric orbit lengths in all networks at different aggregations, from the original version $\phi$ to the fully unweighted simplification $\Phi_{\infty}$. Right: number of orbits of given length in log-log scale observed in food webs in respective aggregations with power-law fits (shown if significant), from $\Phi_1$ (violet) to $\Phi_{\infty}$ (light green).}
   
    \label{fig:OrbitLength}
\end{figure}

Most of the orbits in our dataset contain two or three vertices. Even when considering higher aggregations, orbits rarely contain more than four vertices, as shown in Fig.~\ref{fig:OrbitLength}. This indicates limited functional redundancy, consistent with ecological theory and competitive exclusion principle that only small subsets of species can interchangeably occupy the same trophic role. In network terms, it suggests automorphism groups are small, confirming that food webs are topologically ‘asymmetric’ despite local equivalences.

The vast majority of orbits appear only once aggregation $\Phi_1$ is applied. It indicates that small measurement/estimate differences (smaller than an order of magnitude) indeed mask the approximate structural symmetries. 

Even though $\Phi_2$ aggregations contain more than half of two-element orbits present in $\Phi_{\infty}$, they reveal a much smaller fraction of those containing three and more vertices. This reflects the growing number of conditions that simultaneously have to be fulfilled in larger orbits. If an orbit contains $k$ vertices, there may be $\mathcal{O}(k^2)$ connections between them and $\mathcal{O}(k(N-k))$ connections with other vertices that are all involved in equality conditions.

The analysis of aggregation family $\Phi_{\alpha}$ reveals that statistical trends of orbit lengths inferred for fully unweighted food webs differ from those at $\alpha \neq \infty$. In Fig.~\ref{fig:OrbitLength} we illustrate this by fitting power laws $k^a$ to counts of orbits of different lengths separately for each $\alpha$. Even though they hint at a trend, a pairwise bootstrap procedure confirmed that their differences are not statistically significant for so few points. One can expect nevertheless, that moving from weight-neglecting analyses towards more faithful representations of interaction strengths can alter the inferred distribution of orbit lengths, and that different orbit lengths are affected to different degrees.

These patterns demonstrate that the weight-approximation path $\Phi_{\alpha}$ acts as a continuity bridge between symmetries of the original network and those in its unweighted version. It reveals the level of numerical precision at which functional redundancy becomes structurally manifest and offers a step from qualitative to quantitative studies.

\subsubsection{Trophic Properties}

We analyse two aspects of the trophic structure in relation to symmetric orbits: the trophic span and the mean trophic level.

\begin{figure}[H]
    \centering
    \subfloat[]{
	\includegraphics[width=0.5\linewidth]{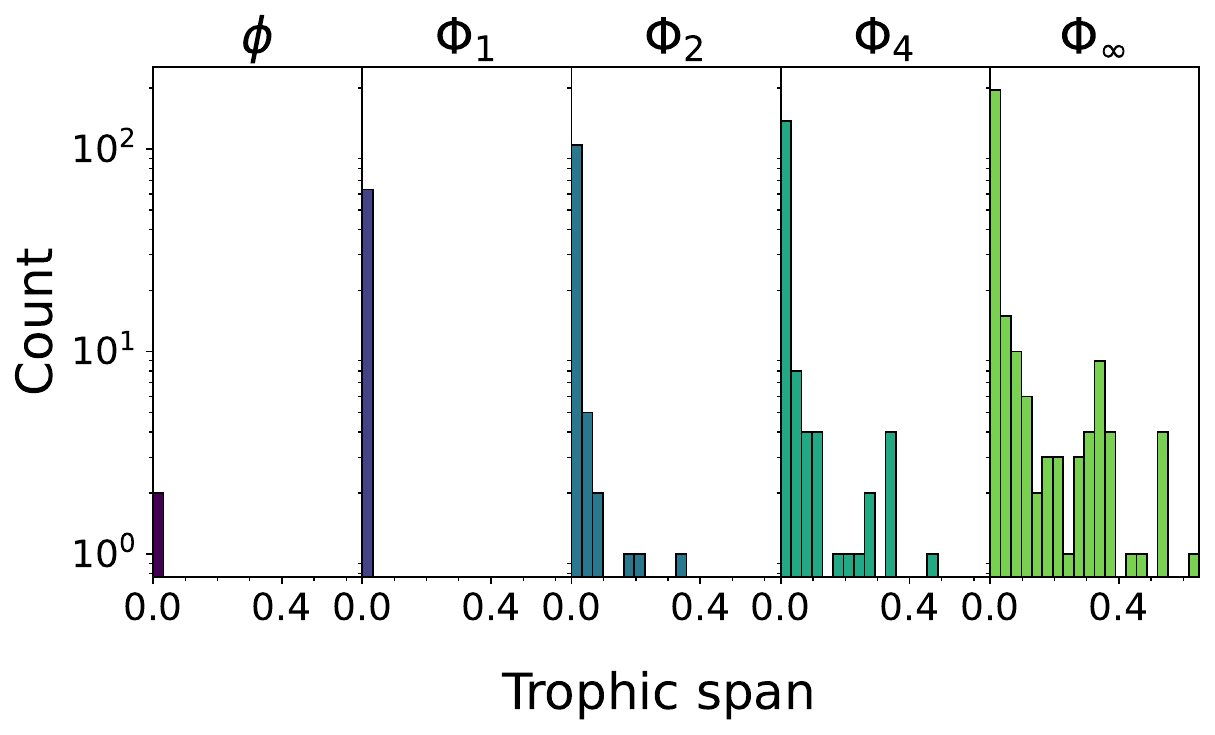}}
    \subfloat[]{
	\includegraphics[width=0.5\linewidth]{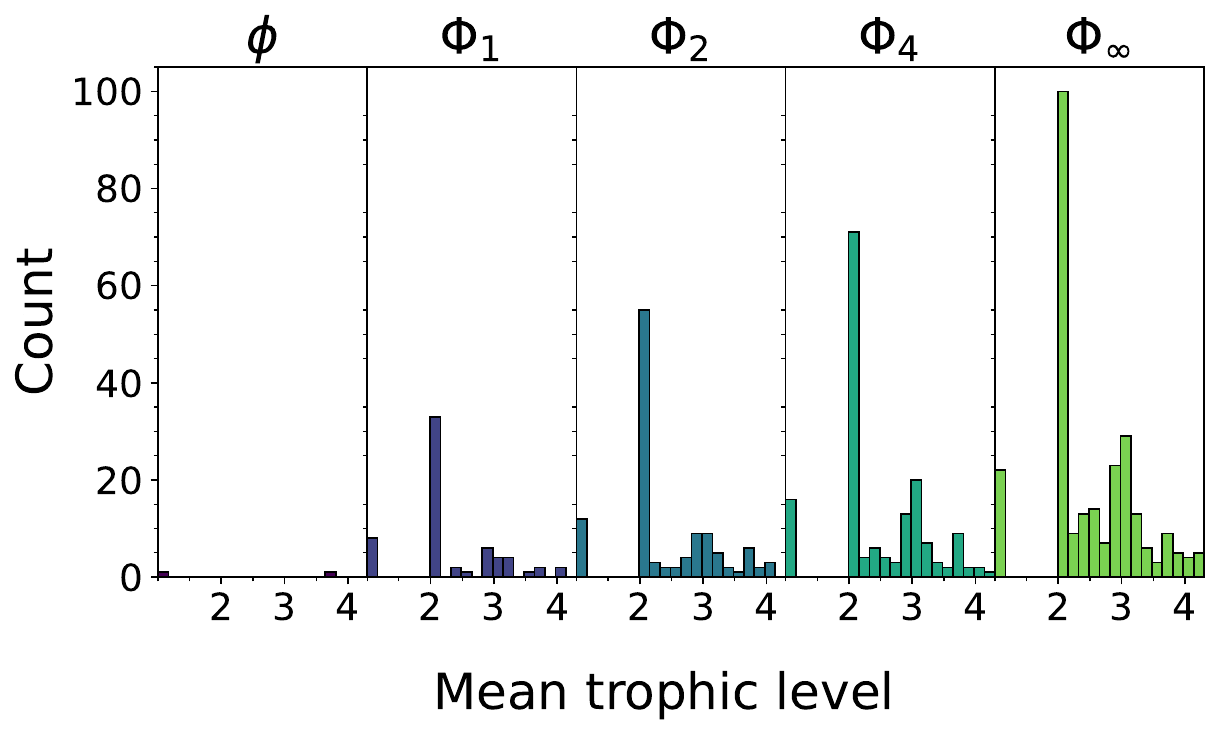}}
    
    \caption{Symmetric orbits can be found at all trophic levels, but especially in the middle ones. The more we approximate edge weights, the larger the possible difference in trophic levels between symmetric vertices becomes. Left: the histogram of symmetric orbits trophic span, in logarithmic scale. Right: the histogram of symmetric orbits mean trophic level.}
    \label{fig: OrbitTrophic}
\end{figure}

The trophic span of an orbit is defined as the difference between the maximum and minimum trophic level of the orbit's vertices. When no aggregation is used, vertices can only be symmetric if they have exactly the same trophic level. Fig.~\ref{fig: OrbitTrophic} (left) shows that in the approximations $\Phi_1, \Phi_2, \Phi_4$ the trophic span remains below 0.5.  In contrast, $\Phi_\infty$ which ignores edge weights, reveals that similarity in ecosystem functions can relate groups at more distant trophic levels. This shows that while qualitative studies might conclude strong competition among differing trophic levels, quantitative comparisons restrict them to practically the same trophic level.

In terms of mean trophic level, symmetric vertices within the same orbit are typically found at trophic levels two or three (see Fig.~\ref{fig: OrbitTrophic}, right). This range corresponds to the highly connected mid-trophic layers, where many interactions concentrate. Although one might expect high connectivity to imply structural asymmetry, we observe no such pattern.

\subsection{Symmetry measures, network size and aggregation}
Symmetry is common in transformed food webs, but almost absent in the originals. Only two untransformed networks contain any orbits, likely due to modeller choices. Non-trivial orbits appear in 36 $\Phi_1$ aggregations and 57 $\Phi_2$ aggregations, corresponding to $14.4\%$ and $22.8\%$, respectively. Their number grows up to 96 networks with non-trivial orbits in unweighted versions, representing $38.4\%$ of the dataset.

\begin{figure}[htbp]
    \centering
    \subfloat[]{\includegraphics[ width = 0.46\textwidth]{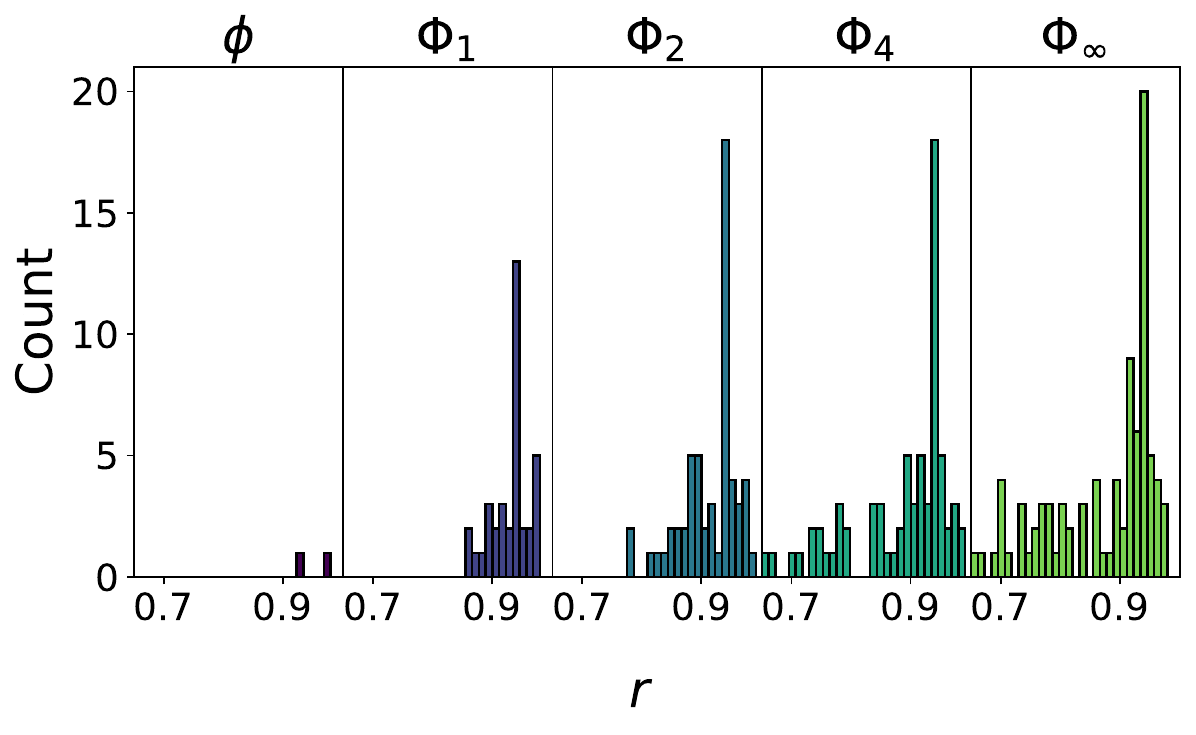}}
    \subfloat[]{\includegraphics[ width = 0.5\textwidth]{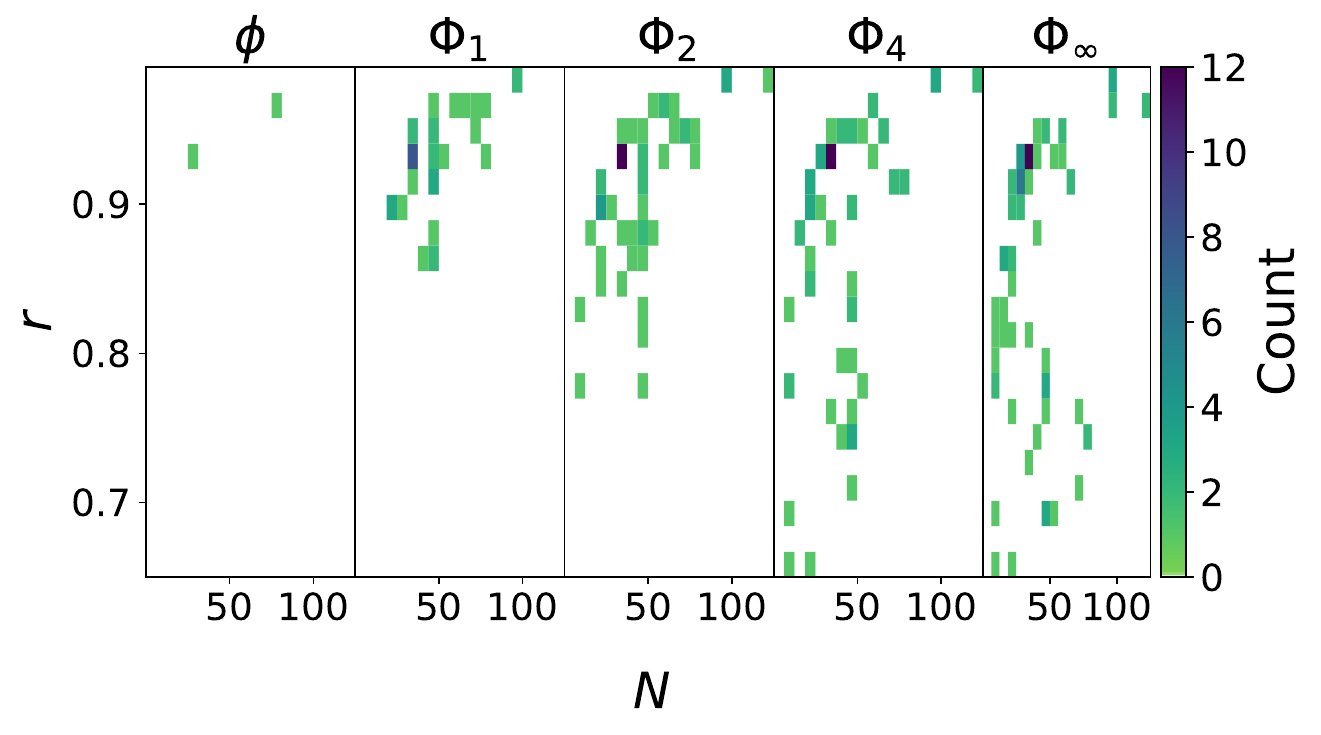}}
    \par
    \subfloat[]{\includegraphics[ width = 0.46\textwidth]{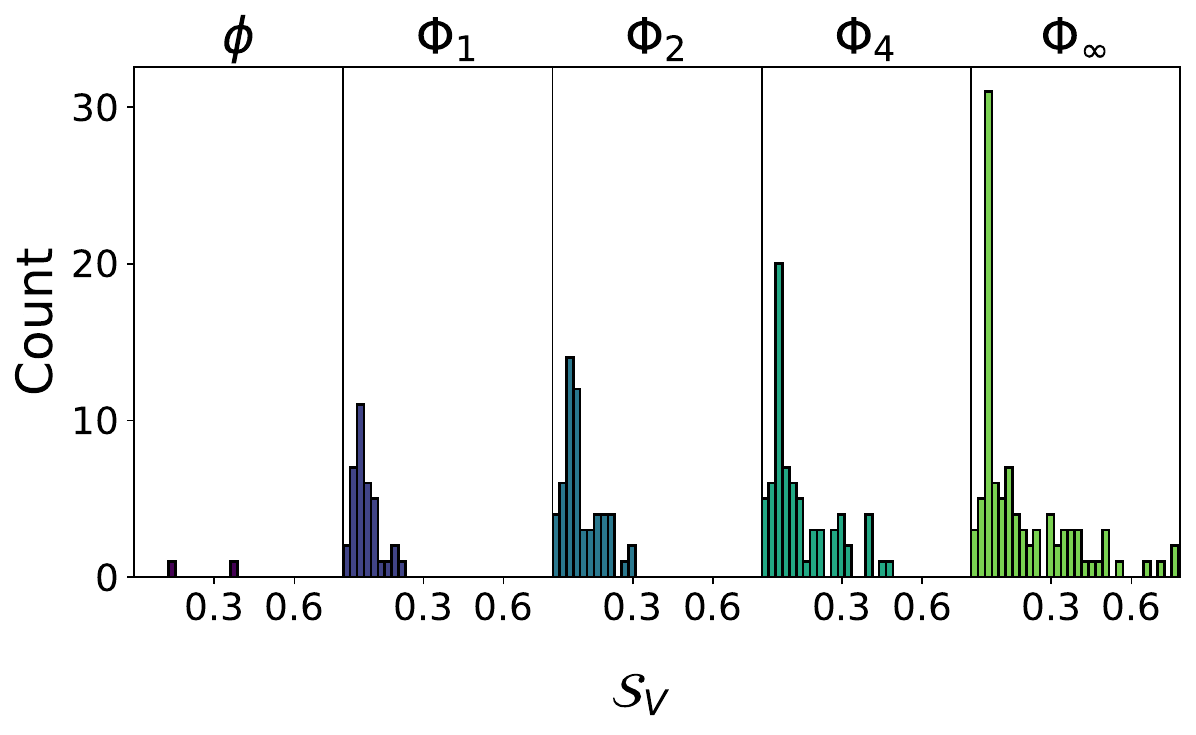}}
    \subfloat[]{\includegraphics[ width = 0.5\textwidth]{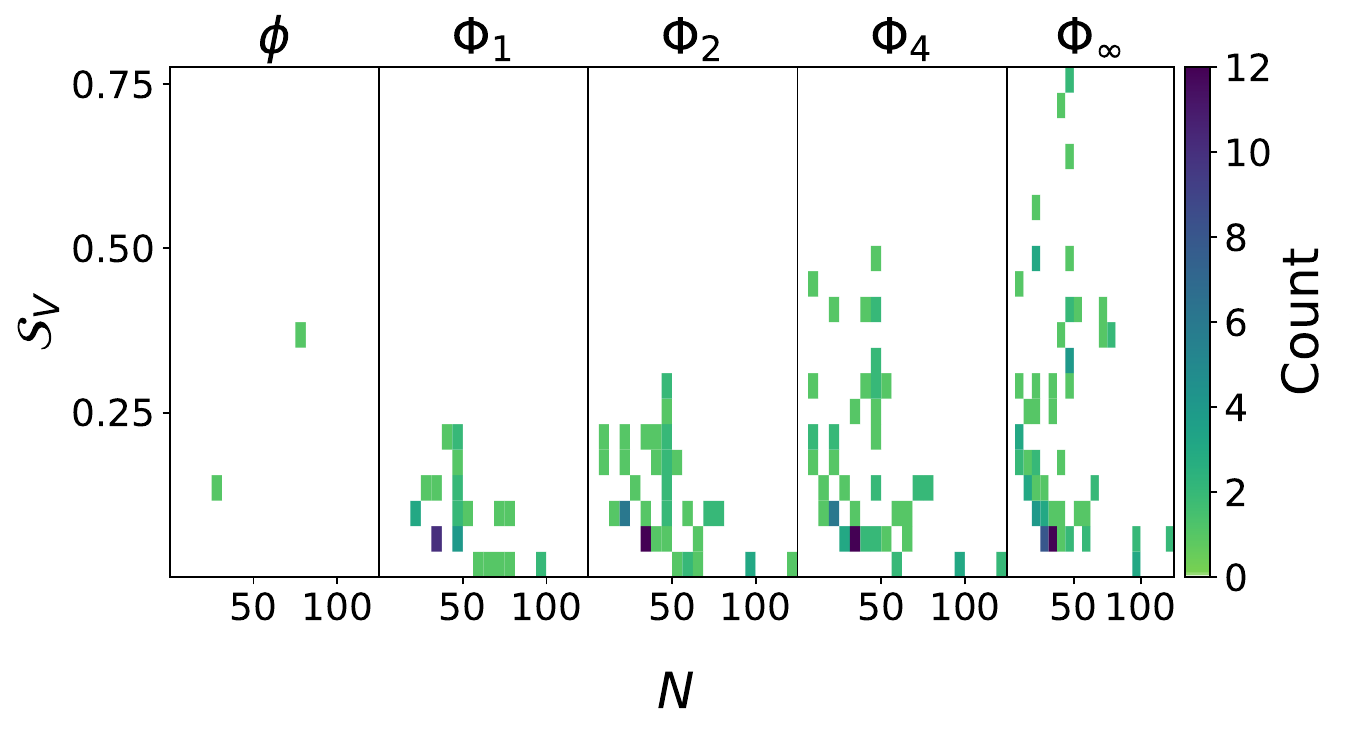}}
    \par
    \subfloat[]{\includegraphics[ width = 0.46\textwidth]{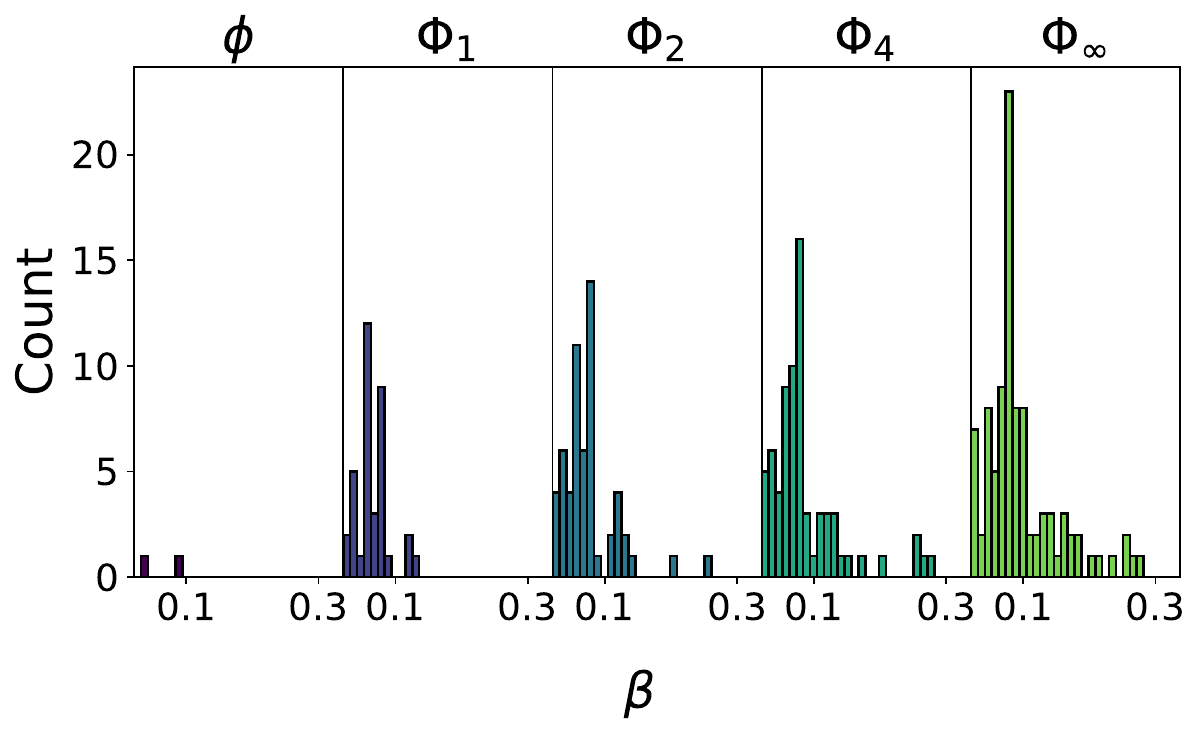}}
    \subfloat[]{\includegraphics[ width = 0.5\textwidth]{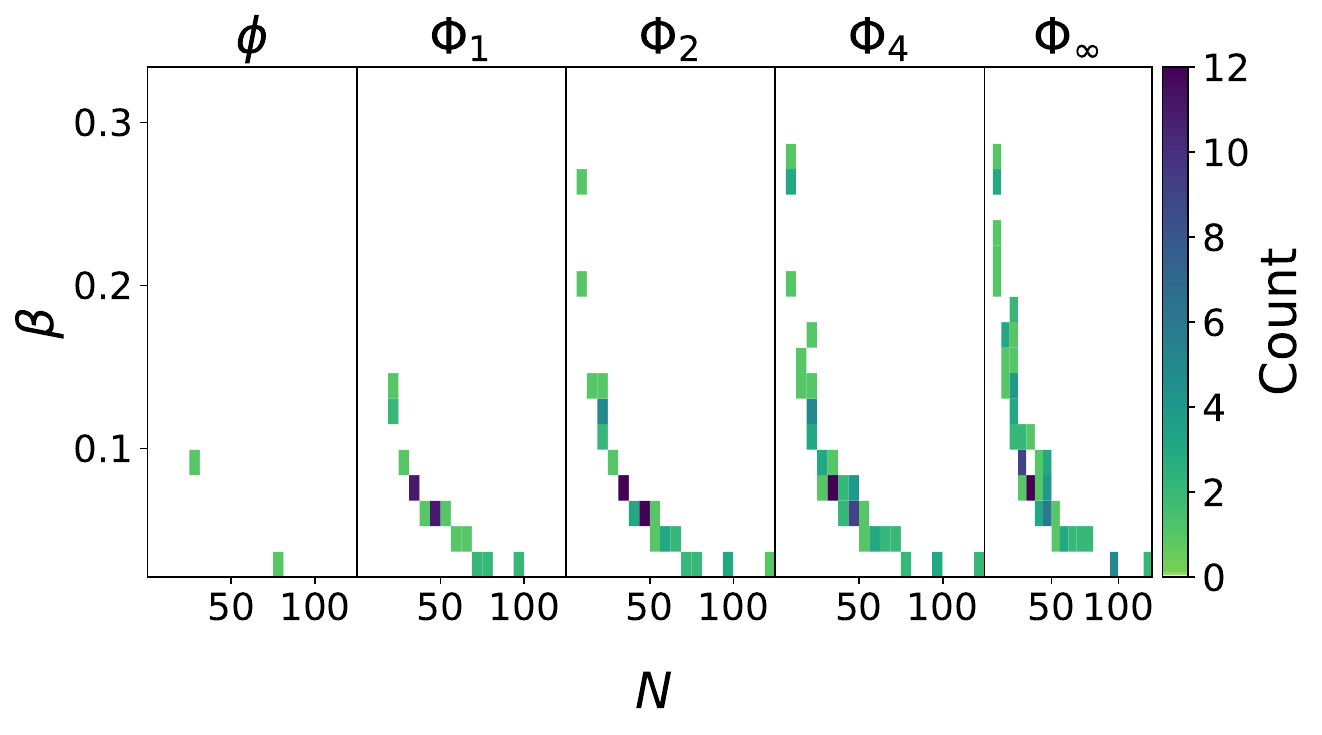}}
    \caption{Left: histograms of symmetry measures, where colour maps the aggregation level. Right: two-dimensional histograms showing the relationship between symmetry measures and network size. The colour of each square represents the number of networks it contains. In all plots, fully asymmetric food webs are omitted.}
    \label{fig: hist_measures_vertices}
\end{figure}

Fig.~\ref{fig: hist_measures_vertices} provides a quantitative overview of symmetry measures in our dataset. The most symmetric networks exhibit redundancy values above 0.8, indicating that symmetric vertices group into small orbits. Typically, only a few symmetric vertices are present. However, when weights are ignored (in $\Phi_\infty$), some networks exhibit over 50\% of vertices in non-trivial orbits. As evident in Fig.~\ref{fig: hist_measures_vertices}, all symmetry measures follow a highly skewed distribution.

The right panel of Fig.~\ref{fig: hist_measures_vertices} illustrates the relationship between symmetry measures and network size. Symmetries (resulting in lower $r$, higher $S_V$, and higher $\beta$) occur most 
frequently in smaller graphs. This is expected: in a weighted directed network, every additional vertex multiplies the number of equality conditions that must hold for an automorphism to exist. Because the number 
of such conditions grows as $\mathcal{O}(k^2)$ with the orbit size $k$, larger networks offer fewer opportunities for weight patterns to align sufficiently to permit vertex permutations. Nevertheless, even in the first aggregation stage, some networks with more than 70 vertices exhibit non-trivial orbits, 
showing that approximate symmetries may persist despite high dimensionality when ecological flows are strongly patterned.

Most networks achieve redundancy values above $0.8$, indicating that automorphisms typically partition the vertex set into many orbits of size one and only a few non-trivial ones. Exceptions with lower redundancy are 
most common among smaller networks, but a few larger ones (around $N\approx 75$) also reach values near $0.7$. Smaller networks display a broader range of symmetric-vertex ratios, whereas some larger networks (with over 50 
vertices) still exhibit even more than $40\%$ symmetric vertices. This highlights that symmetry in food webs is governed less by network size than by the underlying combinatorial constraints imposed by trophic structure.

The measure $\beta$ follows nearly identical curves for all studied graphs. Automorphism groups of empirical food webs are typically small, usually generated by only a few 2-cycles. Thus, $\beta$ is dominated by the factorial term $N!$ in its denominator. As a result, $\beta$ varies primarily with network size rather than with the detailed structure of the orbits. Hence, $\beta$ is best suited for comparing networks of similar size, and should be interpreted cautiously across networks differing substantially in $N$.

The summary of correlations between symmetry measures and the size of the networks is provided in the Table~\ref{tbl: Spearmanvertices}.

\begin{table}[h]
    \centering
    \renewcommand{\arraystretch}{1.2} 
    \scalebox{0.8}{\begin{tabular}{l | c !{\color{gray}\vrule} c | c !{\color{gray}\vrule} c | c !{\color{gray}\vrule} c | c !{\color{gray}\vrule} c}
        
         & \multicolumn{2}{c|}{$\Phi_1$} & \multicolumn{2}{c|}{$\Phi_2$} & \multicolumn{2}{c|}{$\Phi_4$} & \multicolumn{2}{c}{$\Phi_{\infty}$} \\
         & $r_s$ & $p_{\nu}$ & $r_s$ & $p_{\nu}$ & $r_s$ & $p_{\nu}$ & $r_s$ & $p_{\nu}$ \\
        \hline  
        $S_V$ vs $N$  & -0.31 & $6.27\times10^{-2}$ & -0.33 & $1.26\times10^{-2}$ & -0.26 & $3.13\times10^{-2}$ & -0.18 & $7.56\times10^{-2}$ \\
        $r$ vs $N$  & 0.51 & $1.49\times10^{-3}$ & 0.51 & $5.14\times10^{-5}$ & 0.41 & $4.39\times10^{-4}$ & 0.26 & $9.53\times10^{-3}$ \\
        $\beta$ vs $N$  & -0.99 & $1.49\times10^{-32}$ & -0.99 & $5.43\times10^{-47}$ & -0.98 & $1.10\times10^{-53}$ & -0.95 & $1.93\times10^{-48}$ \\
    \end{tabular}}
    \caption{The correlation between symmetry measures quantified by the Spearman rank-order correlation coefficient. $N$ is the number of vertices in the graph, $r_s$ is the Spearman correlation coefficient, and $p_v$ is the p-value of the test.}
    \label{tbl: Spearmanvertices}
\end{table}

Table \ref{tbl: Spearmanvertices} shows that, for each measure, the correlation coefficients seem to decrease as aggregation increases. This weakening correlation mirrors the behaviour of orbit structure under aggregation. As Fig.~\ref{fig:OrbitLength} shows, most non-trivial orbits consist of pairs of vertices. In this case, the number of symmetric vertices and the number of non-trivial orbits are directly proportional, and each symmetry reduces the orbit count only marginally. In contrast, under $\Phi_\infty$--where weights no longer constrain automorphisms--larger orbits emerge, allowing multiple vertices to be grouped together. These larger symmetries lead to a more substantial reduction in orbit structure and a weaker dependence of $r$ on $N$.

Redundancy shows a stronger correlation with $N$ than $S_V$ does. For $\Phi_1$ and $\Phi_2$, the Spearman coefficient ($r_s$) is close to $0.5$, decreasing with coarser 
weight approximations and reaching $0.26$ for $\Phi_\infty$. This weakening correlation mirrors the behaviour of orbit structure under aggregation: as Fig.~\ref{fig:OrbitLength} shows, most orbits consist of exactly two vertices. Pairwise symmetries reduce the number of orbits by exactly one per automorphism pair, which limits their influence on redundancy. Only under $\Phi_\infty$--where weights no longer constrain automorphisms--many more larger orbits appear, and these decrease $r_s$ by allowing more substantial consolidation of orbits.

The predominance of orbits of size two implies that the automorphism groups of these networks consist mostly of disjoint transpositions. Algebraically, this means that symmetries operate locally rather than globally: food webs exhibit isolated pairs of structurally equivalent species rather than large subsets that can be permuted freely. Biologically, this reflects the limited extent of functional redundancy in real ecosystems, where species may share roles pairwise but rarely form large, fully substitutable guilds.

High redundancy ($r > 0.8$) with small orbit sizes suggests that the automorphism group decomposes mainly into 2-cycles, indicating isolated pairs of structurally equivalent species rather than large permutable subgroups.

The measure $\beta$ is almost perfectly correlated with network size: $r_s< -0.98$ across all aggregations except $\Phi_\infty$, where it remains near $-0.94$. This confirms that even when symmetric vertices are present, they do not contribute enough automorphisms to counteract the factorial scaling of $N!$. In ecological terms, this means that while local role equivalence is detectable, food webs remain globally asymmetric and far from regular structures, again reinforcing the scarcity of large permutable groups of species.

\subsection{Relations between symmetry measures}

\begin{figure}[H]
    \centering
    \subfloat[]{\includegraphics[ width = 0.5\textwidth]{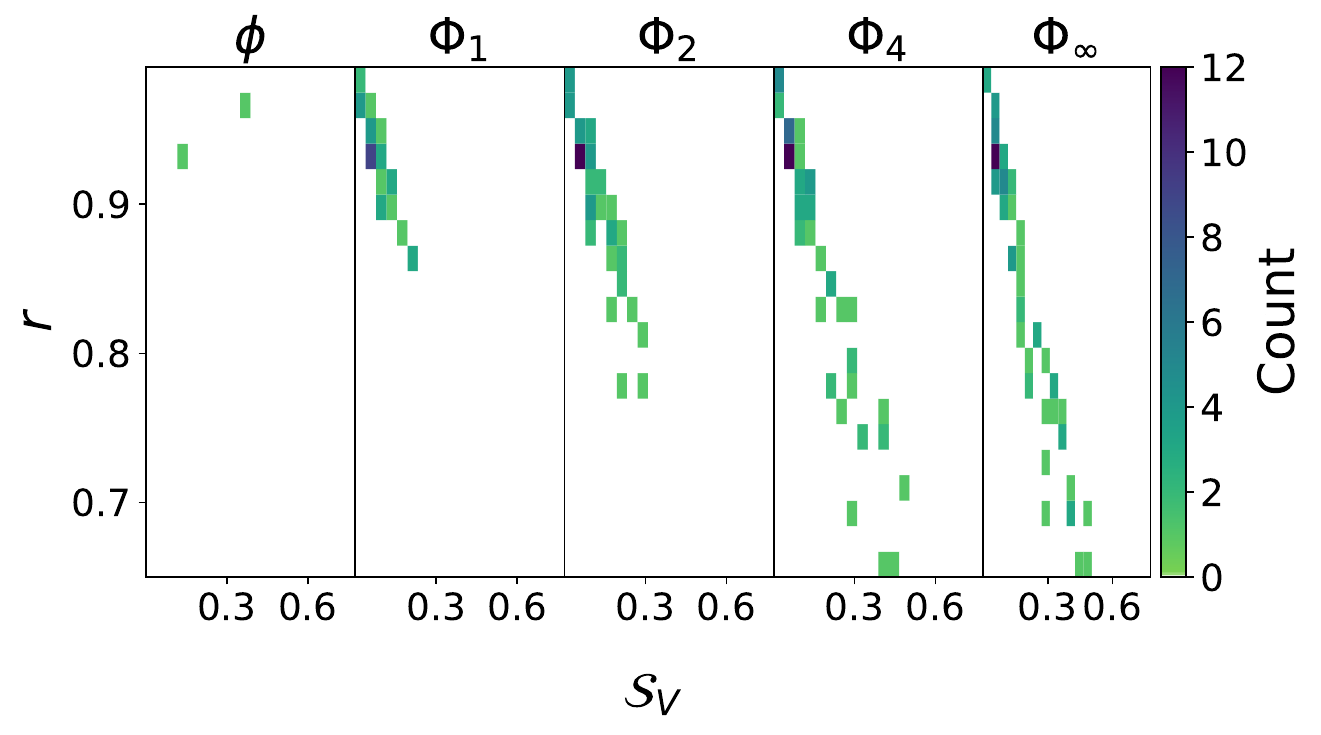}}
    \subfloat[]{\includegraphics[ width = 0.5\textwidth]{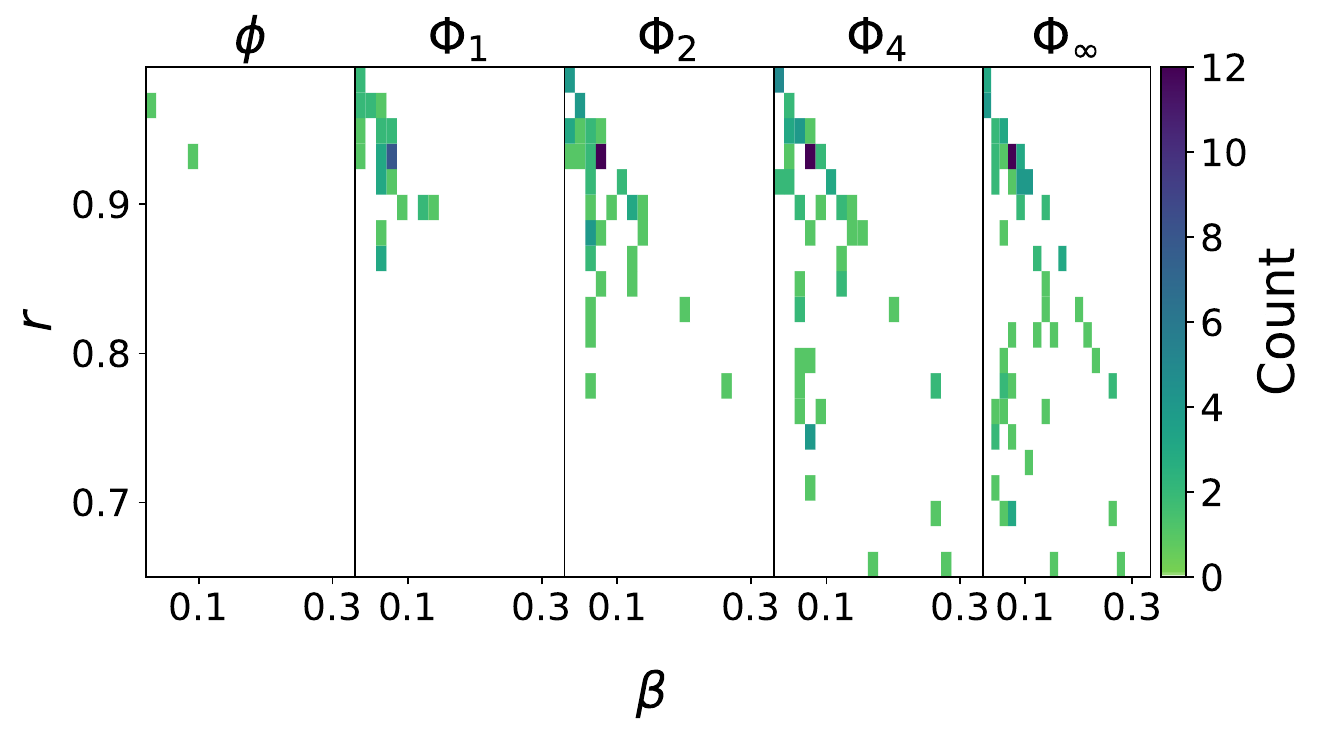}} 
    \par
    \subfloat[]{\includegraphics[ width = 0.5\textwidth]{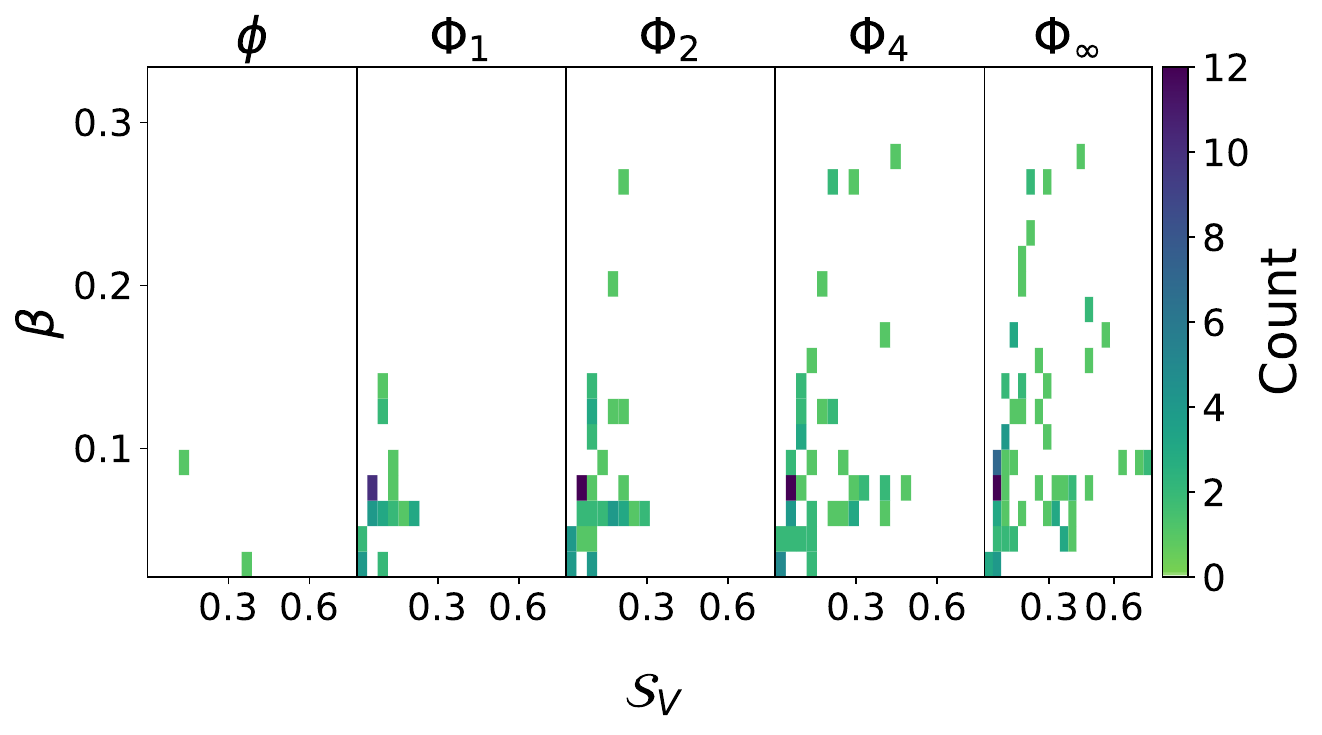}}
    \caption{Two-dimensional histograms showing relations between symmetry measures. The colour of each square represents the number of networks within.}
    \label{fig:2dhist_measures_measures}
\end{figure}

We examine the interdependencies between the vertex, orbit and automorphism aspects of symmetry. Fig.~\ref{fig:2dhist_measures_measures} presents the values of symmetry measure pairs for individual graphs. Table~\ref{tbl: TableSpearmanMeasures} summarises these relationships using Spearman correlation coefficients.

A strong relationship is observed between $S_V$ and $r$, which reflects the dominance of small orbits, particularly those of length two, and pairwise automorphisms. Since most vertices are asymmetric and form orbits of size one, the total number of orbits is high. Merging a few vertices into small symmetric orbits reduces this number only slightly, which explains why redundancy remains above 0.5 even when up to 80\% of vertices belong to symmetric orbits. The Spearman coefficient between these two measures, shown in Table~\ref{tbl: TableSpearmanMeasures}, remains below -0.94 across all levels of aggregation, supporting this observation. This near-perfect correlation between $S_V$ and $r$ implies that most detected symmetries correspond to simple pairwise automorphisms, not more complex subgroup structures.

\begin{table}[h]
    \centering
    \renewcommand{\arraystretch}{1.2} 
    \scalebox{0.84}{\begin{tabular}{l | c !{\color{gray}\vrule} c | c !{\color{gray}\vrule} c | c !{\color{gray}\vrule} c | c !{\color{gray}\vrule} c}
        
         & \multicolumn{2}{c|}{$\Phi_1$} & \multicolumn{2}{c|}{$\Phi_2$} & \multicolumn{2}{c|}{$\Phi_4$} & \multicolumn{2}{c}{$\Phi_{\infty}$} \\
         & $r_s$ & $p_{\nu}$ & $r_s$ & $p_{\nu}$ & $r_s$ & $p_{\nu}$ & $r_s$ & $p_{\nu}$ \\
        \hline  
        $\beta$ vs $r$  & -0.56 & $4.05\times10^{-4}$ & -0.58 & $2.74\times10^{-6}$ & -0.50 & $1.06\times10^{-5}$ & -0.46 & $2.63\times10^{-6}$ \\
        $\beta$ vs $S_V$ & 0.36 & $3.09\times10^{-2}$ & 0.40 & $2.28\times10^{-3}$ & 0.34 & $3.22\times10^{-3}$ & 0.38 & $1.68\times10^{-4}$ \\
        $S_V$ vs $r$  & -0.95 & $2.28\times10^{-18}$ & -0.96 & $8.47\times10^{-33}$ & -0.97 & $8.51\times10^{-45}$ & -0.99 & $6.20\times10^{-74}$ \\
    \end{tabular}}
    \caption{Correlation between symmetry measures based on the Spearman rank-order correlation coefficient. Notation follows Table \ref{tbl: Spearmanvertices}.}
    \label{tbl: TableSpearmanMeasures}
\end{table}

The Spearman coefficient between $\beta$ and $r$ remains close to -0.5 across all aggregations, indicating a weak negative correlation. However, this correlation is stronger than that between $\beta$ and $S_V$. This difference may be due to $\beta$ and $r$ both having a stronger correlation with the number of vertices in the graph.

\section{Discussion}\label{sec:discussion}
Our analysis shows that automorphisms in empirical food webs emerge only when weight heterogeneity is reduced, highlighting the role of the approximation path $\Phi_\alpha$ as a controlled way of revealing near-symmetries. These approximate symmetries correspond to \emph{near-role equivalence}: species that are not strictly automorphic in the original weighted graph but become so once small weight differences are discounted. This interpretation connects automorphism-based symmetry directly to ecological role similarity, offering a mathematically rigorous complement to classical notions such as trophic species or functional groups.

In classical food-web analysis, trophic species are defined by identical sets of predators and prey~\cite{Cohen1990}, and more generally by clustering approaches based on interaction similarity. Standard food-web analyses relied on metrics such as degree, connectance, and interaction similarity~\cite{Dunne2002}. Our approach differs in that it identifies structural equivalence at the level of the entire network, rather than relying on pairwise similarity or heuristic grouping. As such, it provides a complementary, symmetry-based definition of ecological roles.

Automorphism-based role similarity provides a structural perspective also on ecological concepts such as functional redundancy, niche overlap, and response diversity. Vertices belonging to the same orbit are interchangeable within the network structure, suggesting a form of potential functional substitutability analogous to functional redundancy \cite{Walker1992,Petchey2006}. However, unlike standard ecological measures, this notion is derived purely from interaction topology (and discretised flow magnitudes), without requiring species traits or external classifications. 

In contrast to niche overlap, which typically compares similarity of resource use or interaction partners, orbit equivalence captures a stricter, system-wide symmetry that accounts for the entire pattern of interactions. Moreover, the emergence or disappearance of such symmetries under different weight aggregations can be related to response diversity \cite{Elmqvist2003}: species whose structural similarity persists across aggregation levels may represent robust functional groupings, whereas those whose similarity depends on fine-scale weight distinctions may play more differentiated roles. This perspective complements existing ecological measures by identifying structurally defined roles intrinsic to the network representation.

Orbit structure complements standard graph metrics by identifying vertices that are interchangeable within the global network structure, rather than merely similar in local connectivity. A motif-based approach by~\cite{Stouffer2012} offered an intermediate step--using positions occupied by species within smaller subgraphs of food webs to compare species importance. All these complementary efforts help us to understand and structurally define functional roles of species in ecosystems.

Automorphism analysis allows to verify whether redundancy in a network has a more complex pattern or one detectable by investigating neighbourhoods of vertices alone. The dominance of small orbits and high redundancy values across the dataset indicates that automorphism groups are composed mainly of a few independent 2-cycles. This reveals that functional redundancy exists primarily at the level of isolated species pairs rather than larger interchangeable guilds. This observation is consistent with previous findings that empirical food webs exhibit limited large-scale redundancy and are dominated by heterogeneous interaction patterns. Rather than forming large interchangeable trophic guilds, species tend to occupy distinct positions with only occasional local similarities. Our results provide a structural explanation for this pattern, showing that even when approximate symmetries are allowed, they all correspond to independent permutations of mostly small orbits.

Approximate-symmetry frameworks~\cite{Mitra_2006, Pakdemirli_2004, Liu_2020, Pidnebesna_2025, Rosell-Tarrago_2021} typically allow mismatches in topology, such as missing or altered edges. In contrast, food webs require preserving low-weight interactions because weak flows play important roles in system stability~\cite{McCann1998, Jacquet2016}. The weight-approximation strategy used here respects this constraint: it relaxes numerical precision without discarding biologically meaningful links. Combining this approach with selective edge exclusion may offer a promising hybrid method for future research.

Although we focused on logarithmic aggregation--appropriate for the log-normal distribution of biomass flows--other aggregation families are possible. Different functional forms of $\Phi$ would generate different paths through the space of discretisations, potentially uncovering distinct patterns in orbit emergence from original network and coalescence towards those in its unweighted version. In general, the ambiguity introduced by the nature of discretisation of continuous weight distributions will always remain. Observing which subgroups of the $\Phi_{\infty}$ automorphism group appear in how many aggregation schemes may strengthen the corresponding notions of who is more similar to whom in systematic reviews of aggregation functions applied to a given network.

A systematic exploration of aggregation functions could provide deeper insight into the sensitivity of network symmetries to the choice of approximation scale. As every empirical data is approximate due its non-zero uncertainty and modeller choices, it would also offer insights into consequences of such implicit approximations.

Even though the analysed empirical food webs represent various ecosystem types--e.g. continental shelves, bays, estuaries, beaches--the observed rough patterns of symmetries have not reflected significant differences among them.

Finally, we note that food-web models inherently simplify biological reality, and some observed symmetries likely arise from modelling conventions or aggregation decisions. Nonetheless, the minimum approximation level at which two species enter a shared orbit provides a natural, structurally grounded definition of functional similarity that complements trait-based and interaction-based approaches commonly used in food-web analysis.

The dependence of orbit structure on the aggregation scheme has two fundamental sources: the notion of similarity between edge weights encoded by the aggregation function, and discretisation effects arising from partitioning continuous weights into finite categories. In practice, the aggregation should reflect both the distribution and the uncertainty of the data. For food webs, where interaction strengths span several orders of magnitude, logarithmic aggregation provides a natural and interpretable choice. More generally, we recommend comparing symmetries across multiple aggregation schemes and treating as structurally meaningful those that persist across a range of parameter values. In this study, we interpret species similarity specifically under order-of-magnitude approximations of biomass flows.

\section{Conclusions}\label{sec:conclusions}

We have introduced a general framework for detecting approximate symmetries in weighted networks by aggregating edge weights into nested discretisations. This extends automorphism-based analysis from unweighted to weighted graphs while retaining the interpretability of orbit decompositions and efficiency of existing algorithms of automorphism analysis. In food webs, where biomass flows span several orders of magnitude, logarithmic aggregation provides a natural approximation path that reveals the scale at which small weight differences cease to obscure structural similarity.

The case study of the Peruvian upwelling food web shows that species lacking exact symmetries in the original weighted network become automorphic once minor weight discrepancies are relaxed. These orbits represent \emph{near-equivalent} species that occupy nearly identical structural positions and, in several cases, reflect ecological traits---such as body size or trophic niche---not explicitly encoded in the network. This demonstrates that automorphism-based methods can recover latent ecological structure from interaction patterns alone.

Across the full dataset of 250 empirical food webs, the three symmetry measures (symmetric-vertex ratio, redundancy, and the $\beta$ measure) provide complementary insight into the emergence and localisation of network symmetries. Orbit sizes rarely exceed two or three vertices, reflecting the combinatorial fragility of larger symmetric sets in weighted directed graphs. Automorphism equivalence in the analysed networks reduces to structural equivalence, confirming the effectiveness of niche and trophic species concepts used in the ecological literature. Automorphism groups are dominated by isolated transpositions, producing high redundancy and a strong association between redundancy and the fraction of symmetric vertices. Yet symmetric vertices occur in diverse structural positions throughout the networks, and high connectivity does not preclude symmetry, indicating that structural equivalence is not restricted to peripheral or low-degree species. Under the coarsest aggregation, orbits may also connect more distant trophic positions, revealing broader structural analogies in ecosystem organisation.

Overall, the approximation framework provides a principled and interpretable approach for identifying functional similarity in weighted networks. The minimal aggregation level at which two species become substitutable yields a 
quantitative, automorphism-based measure of role similarity, offering a structural perspective on trophic redundancy, niche proximity, and substitutability across ecological communities.

\bibliographystyle{naturemag} 
\bibliography{scibib}

@ARTICLE{Pidnebesna_2025,title={Computing Approximate Global Symmetry of Complex Networks with Application to Brain Lateral Symmetry},year={2025},author={Anna Pidnebesna and David Hartman and Aneta Pokorná and Matěj Straka and Jaroslav Hlinka},doi={10.1007/s10796-025-10585-3},pmid={null},pmcid={null},mag_id={null},journal={Information Systems Frontiers}}

@misc{Ecobase, series={Faculty Research and Publications}, title={EcoBase: A Repository Solution to Gather and Communicate Information from EwE Models},  
DOI={http://dx.doi.org/10.14288/1.0354309}, publisher={University of British Columbia. Fisheries Centre}, author={Coll\'eter, Mathieu and Valls, Audrey Emilie and Guitton, J\'er\^{o}me and Morissette, Lyne and Arregu\'in-S\'anchez, Francisco Francisco and Christensen, Villy and Gascuel, D. (Didier) and Pauly, D. (Daniel)}, year={2013}, collection={Faculty Research and Publications}}

@article{EVERETT1985353,
title = {Role similarity and complexity in social networks},
journal = {Social Networks},
volume = {7},
number = {4},
pages = {353-359},
year = {1985},
issn = {0378-8733},
doi = {https://doi.org/10.1016/0378-8733(85)90013-9},
author = {Martin G. Everett}
}

@ARTICLE{Liu_2020,title={Approximate Network Symmetry},year={2012},author={Yanchen Liu},doi={
10.48550/arXiv.2012.05129},pmid={null},pmcid={null},mag_id={3112882626},journal={arXiv: Physics and Society}}

@article{automorph_distance,
author = {Martínez, Víctor and Berzal, Fernando and Cubero, Juan-Carlos},
title = {An Automorphic Distance Metric and Its Application to Node Embedding for Role Mining},
journal = {Complexity},
volume = {2021},
number = {1},
pages = {5571006},
doi = {https://doi.org/10.1155/2021/5571006},
year = {2021}
}

@book{Albatross,
editor = {Okey, Thomas},
year = {2006},
pages={148},
title = {A trophodynamic model of Albatross Bay, Gulf of Carpentaria: revealing a plausible fishing explanation for prawn catch declines},
publisher={CSIRO Marine and Atmospheric Research Paper 010, Cleveland, Qld, Australia, 148 pp.},
doi={https://doi.org/10.4225/08/5858239b3a821}
}

@ARTICLE{Chesson_2000,title={Mechanisms of Maintenance of Species Diversity},year={2000},author={Peter Chesson },doi={10.1146/annurev.ecolsys.31.1.343},pmid={null},pmcid={null},mag_id={2150597302},journal={Annual Review of Ecology, Evolution, and Systematics}}

@ARTICLE{Chesson_2018,title={Updates on mechanisms of maintenance of species diversity},year={2018},author={Peter Chesson},doi={10.1111/1365-2745.13035},pmid={null},pmcid={null},mag_id={2885393394},journal={Journal of Ecology}}

@ARTICLE{Levine_2017,title={Beyond pairwise mechanisms of species coexistence in complex communities},year={2017},author={Jonathan M. Levine and Jordi Bascompte and Jordi Bascompte and Frederick R. Adler and Peter B. Adler and Stefano Allesina},doi={10.1038/nature22898},pmid={28569813},pmcid={null},mag_id={2616923088},journal={Nature}}

@article{
Pianka_niche_overlap,
author = {Eric R. Pianka },
title = {Niche Overlap and Diffuse Competition},
journal = {Proceedings of the National Academy of Sciences},
volume = {71},
number = {5},
pages = {2141-2145},
year = {1974},
doi = {10.1073/pnas.71.5.2141},
}

@article{Macarthur_realised_niche_1967,
  title={The Limiting Similarity, Convergence, and Divergence of Coexisting Species},
  author={Robert H. Macarthur and Richard A. Levins},
  journal={The American Naturalist},
  year={1967},
  volume={101},
  pages={377 - 385}
}

@book{Elton,
	title = {Animal ecology},
	publisher = {New York, Macmillan Co},
	author = {Elton, Charles S.},
	year = {1927},
	pages = {256},
}

@article{Elmqvist2003,
author = {Elmqvist, Thomas and Folke, Carl and Nyström, Magnus and Peterson, Garry and Bengtsson, Jan and Walker, Brian and Norberg, Jon},
title = {Response diversity, ecosystem change, and resilience},
journal = {Frontiers in Ecology and the Environment},
volume = {1},
number = {9},
pages = {488-494},
doi = {https://doi.org/10.1890/1540-9295(2003)001[0488:RDECAR]2.0.CO;2},
year = {2003}
}

@article{Petchey2006,
author = {Petchey, Owen L. and Gaston, Kevin J.},
title = {Functional diversity: back to basics and looking forward},
journal = {Ecology Letters},
volume = {9},
number = {6},
pages = {741-758},
doi = {https://doi.org/10.1111/j.1461-0248.2006.00924.x},
year = {2006}
}

@book{
Cohen1990,
title = {Community Food Webs},
	publisher = {Springer Berlin, Heidelberg},
	author = {Joel E. Cohen, Frédéric Briand, Charles M. Newman},
	year = {1990},
	pages = {308},
}

@inbook{Mdloti,
author = {Perissinotto, R. and Stretch, Derek and Whitfield, Alan and Adams, Janine and Forbes, Anthony and Demetriades, N.T.},
year = {2011},
month = {01},
pages = {1-69},
chapter = {Ecosystem functioning of temporarily open/closed estuaries in South Africa},
title = {Estuaries: Types, Movement Patterns and Climatical Impacts},
publisher = {Nova Science Publishers}
}

@article{
Bascompte2005,
author = {Jordi Bascompte  and Carlos J. Melián  and Enric Sala },
title = {Interaction strength combinations and the overfishing of a marine food web},
journal = {Proceedings of the National Academy of Sciences},
volume = {102},
number = {15},
pages = {5443-5447},
year = {2005},
doi = {10.1073/pnas.0501562102}}

@article{
Stouffer2012,
author = {Daniel B. Stouffer  and Marta Sales-Pardo  and M. Irmak Sirer  and Jordi Bascompte },
title = {Evolutionary Conservation of Species’ Roles in Food Webs},
journal = {Science},
volume = {335},
number = {6075},
pages = {1489-1492},
year = {2012},
doi = {10.1126/science.1216556},
eprint = {https://www.science.org/doi/pdf/10.1126/science.1216556}}

@article{
Dunne2002,
author = {Jennifer A. Dunne  and Richard J. Williams  and Neo D. Martinez },
title = {Food-web structure and network theory: The role of connectance and size},
journal = {Proceedings of the National Academy of Sciences},
volume = {99},
number = {20},
pages = {12917-12922},
year = {2002},
doi = {10.1073/pnas.192407699},
eprint = {https://www.pnas.org/doi/pdf/10.1073/pnas.192407699}}

@article{Walker1992,
author = {Walker, Brian H.},
title = {Biodiversity and Ecological Redundancy},
journal = {Conservation Biology},
volume = {6},
number = {1},
pages = {18-23},
doi = {https://doi.org/10.1046/j.1523-1739.1992.610018.x},
year = {1992}
}

@manual{sage,
  Key          = {Sage},
  Author       = {William Stein and others},
  Organization = {The Sage Development Team},
  Title        = {{S}age {M}athematics {S}oftware},
  year = {2005-2025},
  note         = {{\tt http://www.sagemath.org}, accessed 2025}
}

@article{Lindeman,
    author = {Lindeman, Raymond L.},
    title = {The trophic-dynamic aspect of ecology},
    journal = {Ecology},
    year = {1942},
    pages = {399–418},
    volume = {23}
}

@article{EwE,
    author = {Villy Christensen, Carl J Walters},
    title = {Ecopath with Ecosim: methods, capabilities and limitations},
    journal = {Ecological Modelling},
    volume = {172},
    year = {2004},
    pages = {109-139},
    issn = {0304-3800},
    doi = {10.1016/j.ecolmodel.2003.09.003}
}

@article{Hutchinson,
    author = {Evelyn G. Hutchinson},
    title = {Concluding remarks},
    journal = {Cold Spring Harbor Symposia on Quantitative Biology},
    year = {1957},
    volume = {22},
    doi = {10.1101/SQB.1957.022.01.039}
}

@book{Pimm1982,
    author = {Stuart L. Pimm},
    title = {Food webs},
    publisher = {Springer Dordrecht},
    year = {1982},
    doi = {10.1007/978-94-009-5925-5}
}

@article{Jacquet2016,
author = {Jacquet, Claire and Moritz, Charlotte and Morissette, Lyne and Legagneux, Pierre and Massol, Franc¸ois and Archambault, Philippe and Gravel, Dominique},
doi = {10.1038/ncomms12573},
isbn = {2041-1723 (Electronic)
2041-1723 (Linking)},
issn = {2041-1723},
journal = {Nature Communications},
pages = {12573},
pmid = {27553393},
title = {No complexity-stability relationship in empirical ecosystems},
volume = {7},
year = {2016}
}

@article{foodwebviz,
author = {Pawluczuk, Lukasz and Iskrzy\'nski, Mateusz},
title = {Food web visualisation: Heat map, interactive graph and animated flow network},
journal = {Methods in Ecology and Evolution},
volume = {14},
number = {1},
pages = {57-64},
doi = {https://doi.org/10.1111/2041-210X.13839},
year = {2023}
}

@article{McCann1998,
author = {McCann, Kevin and Hastings, Alan and Huxel, Gary R.},
doi = {10.1038/27427},
isbn = {0028-0836},
issn = {00280836},
journal = {Nature},
number = {6704},
pages = {794--798},
pmid = {17434536},
title = {{Weak trophic interactions and the balance of nature}},
volume = {395},
year = {1998}
}

@book{Gause1934,
	author = {Gause, Georgii Frantsevich},
	title = {The struggle for existence},
	address = {Baltimore},
	publisher = {Williams \& Wilkins},
	year = {1934}
}

@article{Pauly1998,
  doi = {10.1126/science.279.5352.860},
  year = {1998},
  month = feb,
  publisher = {American Association for the Advancement of Science ({AAAS})},
  volume = {279},
  number = {5352},
  pages = {860--863},
  author = {Daniel Pauly and Villy Christensen and Johanne Dalsgaard and Rainer Froese and Francisco Torres},
  title = {Fishing Down Marine Food Webs},
  journal = {Science}
}

@article{Granovetter_1973,
  author  = {Granovetter, Mark S.},
  title   = {The Strength of Weak Ties},
  journal = {American Journal of Sociology},
  volume  = {78},
  number  = {6},
  pages   = {1360--1380},
  year    = {1973},
  publisher = {University of Chicago Press},
  doi       = {10.1086/225469}
}

@incollection{HollandLeinhardt_1975,
  author    = {Holland, Paul W. and Leinhardt, Samuel},
  title     = {Local Structure in Social Networks},
  booktitle = {Sociological Methodology 1976},
  editor    = {Heise, David R.},
  pages     = {1--45},
  publisher = {Jossey‐Bass},
  address   = {San Francisco},
  year      = {1975}
}

@article{Burt_1976,
  author  = {Burt, Ronald S.},
  title   = {Positions in Networks},
  journal = {Social Forces},
  volume  = {55},
  number  = {1},
  pages   = {93--122},
  year    = {1976},
  publisher = {Oxford University Press}, 
  doi       = {10.2307/2577097}
}

@article{EverettBorgatti_1991,
  author  = {Everett, Martin G. and Borgatti, Steven P.},
  title   = {The Centrality of Groups and Classes},
  journal = {Social Networks},
  volume  = {13},
  number  = {4},
  pages   = {411--426},
  year    = {1991},
  publisher = {Elsevier}
}

@book{Doreian2005, place={Cambridge}, series={Structural Analysis in the Social Sciences}, title={Generalized Blockmodeling}, publisher={Cambridge University Press}, author={Doreian, Patrick and Batagelj, Vladimir and Ferligoj, Anuska}, year={2004}, collection={Structural Analysis in the Social Sciences}}

@ARTICLE{Everett_1994,title={Regular equivalence: general theory},year={1994},author={Martin G. Everett and Martin G. Everett and Stephen P. Borgatti and Stephen P. Borgatti},doi={10.1080/0022250x.1994.9990134},pmid={null},pmcid={null},mag_id={1992028263},journal={Journal of Mathematical Sociology}}

@ARTICLE{Squillace_2025,title={Approximate regular equivalence by partition refinement},year={2025},author={G. Squillace and M. Tribastone and Max Tschaikowski and Andrea Vandin},doi={10.1007/s41109-025-00726-7},pmid={null},pmcid={null},mag_id={null},journal={Applied Network Science}}

@book{Ganter_Wille,
  title={Formale Begriffsanalyse: Mathematische Grundlagen},
  author={Ganter, Bernhard and Wille, Rudolf},
  year={1996},
  publisher={Springer-Verlag},
    doi={https://doi.org/10.1007/978-3-642-61450-7}
}

@article{BORGATTI1993361,
title = {Two algorithms for computing regular equivalence},
journal = {Social Networks},
volume = {15},
number = {4},
pages = {361-376},
year = {1993},
issn = {0378-8733},
doi = {https://doi.org/10.1016/0378-8733(93)90012-A},
author = {Stephen P. Borgatti and Martin G. Everett}
}

@Article{Garlaschelli_review,
AUTHOR = {Garlaschelli, Diego and Ruzzenenti, Franco and Basosi, Riccardo},
TITLE = {Complex Networks and Symmetry I: A Review},
JOURNAL = {Symmetry},
VOLUME = {2},
YEAR = {2010},
NUMBER = {3},
PAGES = {1683--1709},
ISSN = {2073-8994},
DOI = {10.3390/sym2031683}
}

@article{WhiteReitz1983,
title = {Graph and semigroup homomorphisms on networks of relations},
journal = {Social Networks},
volume = {5},
number = {2},
pages = {193-234},
year = {1983},
issn = {0378-8733},
doi = {https://doi.org/10.1016/0378-8733(83)90025-4},
author = {Douglas R. White and Karl P. Reitz}
}

@article{LORRAIN,
title = {Structural equivalence of individuals in social networks},
journal = {Journal of Mathematical Sociology},
volume   = {1},
number   = {1},
pages= {49-80},
doi = {10.1080/0022250X.1971.9989788},
year = {1971},
author = {Francois Lorrain and Harrison C. White}
}

@article{Yodzis1999,
  title={In search of operational trophospecies in a tropical aquatic food web},
  author={Yodzis, P. and Winemiller, K. O.},
  journal={Oikos},
  volume={87},
  number={2},
  pages={327--340},
  year={1999},
  doi={10.2307/3546747}
}

@article{Williams2000,
  title={Simple rules yield complex food webs},
  author={Williams, R. J. and Martinez, N. D.},
  journal={Nature},
  volume={404},
  number={6774},
  pages={180--183},
  year={2000},
  doi={10.1038/35004572}
}

@article{Fulton2003,
  title={Effect of complexity on marine ecosystem models},
  author={Fulton, E. A. and Smith, A. D. M. and Johnson, C. R.},
  journal={Marine Ecology Progress Series},
  volume={253},
  pages={1--16},
  year={2003},
  doi={10.3354/meps253001}
}

@article{Brose2006,
  title={Allometric scaling enhances stability in complex food webs},
  author={Brose, U. and Williams, R. J. and Martinez, N. D.},
  journal={Ecology Letters},
  volume={9},
  number={11},
  pages={1228--1236},
  year={2006},
  doi={10.1111/j.1461-0248.2006.00978.x}
}

@article{Newman2006,
  title={Modularity and community structure in networks},
  author={Newman, M. E. J.},
  journal={Proceedings of the National Academy of Sciences},
  volume={103},
  number={23},
  pages={8577--8582},
  year={2006},
  doi={10.1073/pnas.0601602103}
}

@article{Fortunato2010,
  title={Community detection in graphs},
  author={Fortunato, S.},
  journal={Physics Reports},
  volume={486},
  number={3--5},
  pages={75--174},
  year={2010},
  doi={10.1016/j.physrep.2009.11.002}
}

@article{SanchezGarcia2020,
   title={Exploiting symmetry in network analysis},
   volume={3},
   ISSN={2399-3650},
   DOI={10.1038/s42005-020-0345-z},
   number={1},
   journal={Communications Physics},
   publisher={Springer Science and Business Media LLC},
   author={Sánchez-García, Rubén J.},
   year={2020},
   month={May}
}

@article{MacArthur20083525,
title = {Symmetry in complex networks},
journal = {Discrete Applied Mathematics},
volume = {156},
number = {18},
pages = {3525-3531},
year = {2008},
issn = {0166-218X},
doi = {10.1016/j.dam.2008.04.008},
author = {Ben D. MacArthur and Rubén J. Sánchez-García and James W. Anderson},
}

@article{HolmeDetectingDegree,
  title = {Detecting degree symmetries in networks},
  author = {Holme, Petter},
  journal = {Phys. Rev. E},
  volume = {74},
  issue = {3},
  pages = {036107},
  numpages = {7},
  year = {2006},
  month = {Sep},
  publisher = {American Physical Society},
  doi = {10.1103/PhysRevE.74.036107},
  }

@article{Hu2014,
author = {Hu, Chenhui and Fakhri, Georges and Li, Quanzheng},
year = {2014},
month = {09},
journal = {Medical Image Computing and Computer-Assisted Intervention – MICCAI 2014},
pages = {733-40},
title = {Evaluating Structural Symmetry of Weighted Brain Networks via Graph Matching},
volume = {17},
isbn = {978-3-319-10469-0},
doi = {10.1007/978-3-319-10470-6_101}
}

@article{Xiao2008,
   title={Symmetry-based structure entropy of complex networks},
   volume={387},
   ISSN={0378-4371},
   DOI={10.1016/j.physa.2008.01.027},
   number={11},
   journal={Physica A: Statistical Mechanics and its Applications},
   publisher={Elsevier BV},
   author={Xiao, Yang-Hua and Wu, Wen-Tao and Wang, Hui and Xiong, Momiao and Wang, Wei},
   year={2008},
   month={Apr},
   pages={2611–2619}
}

@ARTICLE{Pakdemirli_2004,title={Comparison of Approximate Symmetry Methods for Differential Equations},year={2004},author={Mehmet Pakdemirli and Muhammet Yürüsoy and İhsan Timuçin Dolapçı},doi={10.1023/b:acap.0000018792.87732.25},pmid={null},pmcid={null},mag_id={2008884333},journal={Acta Applicandae Mathematicae}}

@ARTICLE{Mitra_2006,title={Partial and approximate symmetry detection for 3D geometry},year={2006},author={Niloy J. Mitra and Leonidas J. Guibas and Leonidas J. Guibas and Mark Pauly and Mark Pauly},doi={10.1145/1141911.1141924},pmid={null},pmcid={null},mag_id={2060206980},journal={ACM Transactions on Graphics}}

@ARTICLE{Rosell-Tarrago_2021,title={Quasi-symmetries in complex networks: a dynamical model approach},year={2021},author={Gemma Rosell-Tarragó and Albert Díaz-Guilera},doi={10.1093/comnet/cnab025},pmid={null},pmcid={null},mag_id={3197243492},journal={Journal of Complex Networks}}

@inproceedings{Peru,
author={A. Jarre-Teichmann and D. Pauly},
title={Seasonal changes in the Peruvian upwelling ecosystem},
year={1993},
pages={307-314},
editor={ V. Christensen and D. Pauly},
booktitle={Trophic models of aquatic ecosystems. ICLARM Conference Proceedings},
volume={26}
}

@article{MacArthurAndersonBeta,
author = {Ben D. MacArthur and James W. Anderson},
title = {Symmetry and Self-Organization in Complex Systems},
year = {2006},
journal={arXiv},
doi = {10.48550/arXiv.cond-mat/0609274},
}

\end{document}